\documentclass{lmcs} 
\pdfoutput=1
\usepackage[utf8]{inputenc}

\usepackage{lastpage}
\lmcsdoi{21}{4}{11}
\lmcsheading{}{\pageref{LastPage}}{}{}%
{Sep.~22,~2023}{Oct.~17,~2025}{}

\usepackage{cite}

\usepackage{amsmath,amssymb,amsfonts}
\usepackage{algorithmic}
\usepackage{graphicx}
\usepackage{textcomp}

\usepackage{boondox-cal}

\usepackage{arydshln}

\usepackage{subfig}

\newcommand{\ol}[1]{\overline{#1}}
\newcommand{\ul}[1]{\underline{#1}}

\newcommand{\rb}[1]{\left({#1}\right)}

\newcommand{\sqb}[1]{\left[{#1}\right]}

\newcommand{\sqbs}[1]{[{#1}]}

\newcommand{\intv}[2]{\left[{#1}\; ; \; {#2}\right]}

\newcommand{\dbr}[1]{\langle {#1} \rangle}

\keywords{Interval analysis, uncertain systems, parameter identification, state estimation, guaranteed stabilizing control, model predictive control, interval observers}

\usepackage{hyperref}
\theoremstyle{plain} 

\begin{document}

\title[Offline and Online Use of Interval and Set-Based Approaches]{Offline and Online Use of Interval and Set-Based Approaches for Control and State Estimation: A {Selection} of Methodological Approaches and Their Application}

\author[A.~Rauh]{Andreas Rauh\lmcsorcid{0000-0002-1548-6547}}[a]
\author[M.~Lahme]{Marit Lahme\lmcsorcid{0000-0001-8633-1908}}[a]
\author[S.~Rohou]{Simon Rohou\lmcsorcid{0000-0001-6232-9918}}[b]
\author[L.~Jaulin]{Luc Jaulin\lmcsorcid{0000-0002-0938-0615}}[b]
\author[T.N.~Dinh]{Thach Ngoc Dinh\lmcsorcid{0000-0001-8827-0993}}[c]
\author[T.~Ra\"issi]{Tarek Ra\"issi\lmcsorcid{0000-0002-8735-9395}}[c]
\author[M.~Fnadi]{Mohamed Fnadi\lmcsorcid{0000-0001-9593-8859}}[d]

\address{Distributed Control in Interconnected Systems, Department of Computing Science, Carl von Ossietzky Universität Oldenburg, D-26111 Oldenburg, Germany}	
\email{andreas.rauh@uni-oldenburg.de, marit.lahme@uni-oldenburg.de}  

\address{ENSTA, Lab-STICC (Robex), F-29806 Brest, France}	
\email{simon.rohou@ensta.fr, luc.jaulin@ensta.fr}  

\address{Cedric-Laetitia, Conservatoire National des Arts et Métiers, 
F-75141 Paris, France}	
\email{ngoc-thach.dinh@lecnam.net, tarek.raissi@cnam.fr}  

\address{Laboratoire d'Informatique, Signal et Image de la Côte d'Opale, Université du Littoral Côte d'Opale, F-62228, France}	
\email{mohamed.fnadi@univ-littoral.fr}  





\begin{abstract}
  \noindent Control and state estimation procedures need to be robust against imprecisely known parameters, uncertainty in initial conditions, and external disturbances. Interval methods and other set-based techniques form the basis for the implementation of powerful approaches that can be used to identify parameters of dynamic system models in the presence of the aforementioned types of uncertainty. Moreover, they are applicable to a verified feasibility and stability analysis of controllers and state estimators. In addition to these approaches {which are typically used offline} for analysis {of system models designed with classical floating point procedures}, interval and set-based methods have also been developed in recent years, which allow to {directly} solve the associated design tasks and to implement reliable techniques that are applicable online{, i.e., during system operation}. The latter approaches include set-based model predictive control, online parameter adaptation techniques for nonlinear variable-structure and backstepping controllers, interval observers, and fault diagnosis techniques. This paper provides an overview of the methodological background and reviews numerous practical applications for which interval and other set-valued approaches have been employed successfully. 
\end{abstract}

\maketitle

\section{Introduction}\label{sec:intro}
Interval analysis has become an active field of research over the past decades. {Especially, the foundations for its modern use} can be traced back to the works of R.E.~Moore and his famous book published in 1966~\cite{Moore:1966}. {However, even earlier evidence from the 19th century can be found if not only works aiming at computer implementations are accounted for but if also foundations of truncated, finite-precision arithmetic are considered~\cite{Kearfott2023,Schnellinger}.}

\subsection{{Use of Interval Methods to Deal with Bounded Uncertainty}}
In its widest sense, interval analysis provides an approach to perform computations on finitely large sets in such a way that the determined solutions contain all the solutions of the mathematically stated problem with certainty~\cite{mayer2017interval,AppliedIntervalAnalysis}. In such a way, it is not only possible to account for numerical errors due to the limited precision of the numbers representable in floating point arithmetic on a CPU, but also for discretization and method errors. The latter arise inevitably when solving, for example, ordinary differential equations with the help of temporal series expansions truncated after a finite order \cite{Ned:2006,Ned:2011,Kapela,Lohner:1987,BUNGER2020112511}. In addition to these effects, which are small (at least) for a single evaluation step, but can cause severe damage if neglected, the same techniques also allow for computing with wider sets that result from imperfectly known system parameters and initial conditions in (technical) applications. This lack of knowledge is commonly referred to as epistemic uncertainty~\cite{KIUREGHIAN2009105}. As such, interval analysis provides a mathematically sound way to handle epistemic uncertainty that can furthermore also {result from} other kinds of a lack of knowledge of the actual system behavior, for example, friction phenomena in dynamics of rigid bodies {or} fluidic applications.

In addition to the application domains mentioned above, further high-level functionalities have been developed on the basis of interval analysis. These contain techniques for solving sets of nonlinear algebraic equations with the help of Newton-type methods \cite{Krawczyk:1969,Neumaier:1990} or the identification of feasible solution sets that satisfy (nonlinear) constraints in a guaranteed manner \cite{Sandretto.2014}. In the latter class of problems, the SIVIA algorithm \cite{AppliedIntervalAnalysis,Walter:1994} (set inversion via interval analysis) as well as contractor approaches~{\cite{Sandretto.2014} serve as the fundamental solution methodologies}, commonly being combined with Newton-type methods. Moreover, extensions exist that allow for handling outliers in a suitable manner by relaxed set intersections \cite{Jaulin.2009}.

\subsection{{Notations}} {Throughout this article, real-valued scalar interval variables are denoted in the form
\begin{equation}
	 x\in \sqb{x} := \intv{\underline{x}}{\overline{x}}
\end{equation}
for which ${\underline{x}}\le{\overline{x}}$ holds. The interval bounds ${\underline{x}},{\overline{x}}\in \mathbb{R}$ are commonly also denoted as the infimum $\inf\rb{\sqb{x}}$ and supremum $\sup\rb{\sqb{x}}$. To distinguish vectors and matrices from scalar quantities, boldface lower- and upper-case letters are used, respectively. Interval vectors and matrices, denoted, for example, by $\sqb{\mathbf{a}}$ and $\sqb{\mathbf{A}}$, are defined in an elementwise manner as the Cartesian product of scalar intervals.}

{For a set-based generalization of the evaluation of a (multi-variate, vector-valued) function $\mathbf{y} = \mathbf{f}\rb{\mathbf{x}} : \mathbb{R}^n \mapsto \mathbb{R}^m$, the so-called interval extension $\sqb{\mathbf{y}} = \sqb{\mathbf{f}}\rb{\sqb{\mathbf{x}}}$ is defined~\cite{AppliedIntervalAnalysis,Moore:1966,mayer2017interval}. It satisfies the property 
\begin{equation}\label{eq:encl:fcn}
	\sqb{\mathbf{f}}\rb{\sqb{\mathbf{x}}} \supseteq
	\left\{
	\mathbf{f} \rb{\mathbf{x}} \mid \mathbf{x} \in \sqb{\rb{\mathbf{x}}}
	\right\}
	\enspace .
\end{equation}
}

\subsection{{Challenges of the Use of Set-Based Methods}}
Besides the {advantageous feature that interval techniques} always provide guaranteed enclosures of a mathematically formulated problem {in the sense~(\ref{eq:encl:fcn})}, they also come with the disadvantage of a certain degree of pessimism. This pessimism is caused by the following two fundamental issues \cite{AppliedIntervalAnalysis,mayer2017interval,Cornelius:1984}:
\begin{itemize}
	\item \emph{The dependency effect:} The difference of two intervals with numerically identical bounds generally does not simplify to the value zero as shown in
	\begin{equation}
		\intv{0}{1} - \intv{0}{1} = \intv{-1}{1} \enspace.
	\end{equation}
	This is due to the fact that the difference in the equation above does not carry any information on whether both arguments are identical quantities or not. Therefore, variables are generally treated as independent in any arithmetic operation according to 
	\begin{equation}
		\left\{ x \circ y | x\in \sqb{x} := \intv{\underline{x}}{\overline{x}}, y\in \sqb{y} := \intv{\underline{y}}{\overline{y}}		
		\right\}~,
		\quad \text{where} \quad 
		\circ \in \{+,-,\cdot, /\}\enspace ,
	\end{equation}
	unless simplifications can be carried out on the basis of symbolic formula manipulation, which allows to exploit knowledge that the two arguments $x$ and $y$ represent exactly the same (physical) quantity. Therefore, in classical interval analysis, there {does not exist any inversion} for the operations of set-based addition and multiplication.
	\item \emph{The wrapping effect:} Transformations of sets, for example, rotations and scalings even in linear matrix-vector products as in the recursive evaluation of discrete-time dynamic systems
	\begin{equation}\label{eq:iter}
		[{\mathbf{x}}](t_{k+1}) = \mathbf{A}\cdot [{\bf{x}}](t_{k})
	\end{equation}
	with the exemplarily chosen initial conditions and dynamics matrix
	\begin{equation}\label{eq:iter1}
		\mathbf{x}(t_0) \in 
		[{\mathbf{x}}](t_{0}) = \begin{bmatrix}\left[-1 \; ; \; 1\right] \\ \left[-1 \; ; \; 1\right] \end{bmatrix},~
		\mathbf{A}  = {\frac{1}{\sqrt{2}}}\begin{bmatrix}1 & 1\\ -1 & 1 \end{bmatrix}
	\end{equation}
	may lead to an exponential increase in the volume of the solution enclosures
	\begin{equation}\label{eq:iter1b}
		\mathbf{x}(t_k) \in 2^\frac{k}{2} \cdot \begin{bmatrix}\left[-1 \; ; \; 1\right] \\ \left[-1 \; ; \; 1\right] \end{bmatrix}
	\end{equation}
	if axis-aligned interval box enclosures were determined in each step~(\ref{eq:iter}). {This growth appears} despite the fact that the true volume would stay constant in the example above because the matrix $\mathbf{A}$ in~(\ref{eq:iter1}) represents a volume-preserving rotation of the initial box by $45^\circ$ in each step. This preservation of volume is especially visible in even sampling steps, where for example the relations
	\begin{equation}
		\mathbf{A}^2  = \begin{bmatrix}0 & 1\\ -1 & 0 \end{bmatrix}
		\quad 
		\text{and}
		\quad 
		\mathbf{A}^4  = {\frac{1}{\sqrt{2}}}\begin{bmatrix}1 & 0\\ 0 & 1 \end{bmatrix}
	\end{equation}
	hold.
\end{itemize}

\subsection{{Selected Countermeasures Against Overestimation}}
{In the example~(\ref{eq:iter})--(\ref{eq:iter1b})}, changing the evaluation order according to
\begin{equation}\label{eq:iter2}
	[{\mathbf{x}}](t_{k+1}) = \mathbf{A}^{k+1} \cdot [{\bf{x}}](t_{0})
	\enspace,
\end{equation}
where the integer-order powers of the point-valued matrix $\mathbf{A}$ are firstly determined, allows to solve this problem \cite{Irina}. However, there is no general solution available, especially in cases where the transformations are nonlinear or the matrices $\mathbf{A}$ may themselves include intervals in each element. 

Then, a large variety of other set-representations may turn out to be more efficient, for example, the use of ellipsoids \cite{Kurzhanskii,9655062} (being mapped exactly to an ellipsoid in point-valued linear transformations), (constrained) zonotopes \cite{ifqir2022zonotopic,REGO2022110425}, zonotope bundles \cite{6160872}, or polytopes \cite{Rauh_Rohou,ROHOU2021104951}. In nonlinear settings, the computationally inexpensive thick ellipsoid approach recently published by two of the authors provides the possibility to detect pessimism in the guaranteed outer enclosure by performing a comparison with an inner ellipsoidal enclosure belonging to the solution set with certainty \cite{a14030088,Rauh_Jaulin}. Moreover, computationally more demanding techniques such as Taylor model arithmetic \cite{BUNGER2020112511,Buenger_2018,Florian_Buenger2021,Hoefkens2} provide further means to reduce the wrapping effect that is {omnipresent in any dynamic system simulation or during finding solutions to algebraic systems of equations.} {Taylor model arithmetic has also been employed in the frame of guaranteed state estimation, especially in predictor--corrector approaches, as well as in the (offline) verification of the robostness of control strategies for nonlinear uncertain systems~\cite{4402415,ANTRITTER2012452,KLETTING2007876}. Although Taylor models allow for effectively reducing the wrapping effect by representing nonlinear dependencies between the individual state variables, they have the following drawbacks in an online use: First, intersections between predicted domains and information obtained by sensors is computationally expensive during state estimation. Second, strategies for limiting the Taylor model order are inevitable to limit and keep the computational complexity constant during online use. For that reasons, zonotopes as well as ellipsoids have shown to represent a good compromise with respect to a slight decrease in the state enclosure quality but a notable reduction in computational effort.}

{As an additional option}, coordinate transformations, as derived in \cite{kersten_rauh:2018,Kersten.2019,MazBern} for systems with either real or complex eigenvalues after their linearization, as well as the exploitation of specific monotonicity properties, allow for reducing both the pessimism discussed above and the computational effort required for its reduction. An example for the exploitation of monotonicity properties is the design of interval observers \cite{tarek_denis1,tarek_denis2} that allows for computing elementwise lower and upper bounds for the state trajectories according to the following scheme \cite{smith}. Here and in the remainder of the article, the word observer denotes an estimation scheme that allows for reconstructing information about not directly measurable internal state variables by means of a model of the considered dynamic system as well as information about its control signals and measured outputs.

An autonomous set of ordinary differential equations 
	\begin{equation}\label{eq:state_general}
		\dot{\mathbf{x}}(t) = {\mathbf{f}}\rb{{\mathbf{x}}(t)}~,~~
		\mathbf{x} \in \mathbb{R}^n
	\end{equation}
	is cooperative according to \cite{Angeli:2003,smith,Hirsch}
	if the property
	\begin{equation}
		x_i^{\dbr{1}}\rb{t}   \ge x_i^{\dbr{2}}\rb{t}
	\end{equation}
	holds at each $t\ge t_0$ for all elements $i \in \{1, \ldots, n\}$ of the two vectors with  $\mathbf{x}^{\dbr{1}}\rb{t_0}$ and $\mathbf{x}^{\dbr{2}}\rb{t_0}$ being the initial conditions which satisfy again the inequalities
	\begin{equation}
		x_i^{\dbr{1}}\rb{t_0} \ge x_i^{\dbr{2}}\rb{t_0}
		\quad \text{for all} \quad i \in \{1, \ldots, n\}
		\enspace .
	\end{equation}
	For continuous-time systems, cooperativity can be checked by the sufficient sign conditions 
	\begin{equation}\label{eq:coop2}
		J_{i,j}\rb{\mathbf{x}} \ge 0~,~~
		i,j\in \{1,\ldots, n\}~,~~i\neq j
	\end{equation} 
	for all off-diagonal elements of the Jacobian 
	\begin{equation}\label{eq:coop1}
		{\mathbf{J}}\rb{\mathbf{x}} = \frac{\partial{\mathbf{f}}}{\partial{\mathbf{x}}}\rb{{\mathbf{x}}}
	\end{equation}
	of the right-hand side of the state equations evaluated for all reachable states ${\mathbf{x}} = {\mathbf{x}}(t)$. Matrices satisfying this non-negativity property for the off-diagonal elements are also denoted as \emph{Metzler} matrices in the literature~\cite{smith,Rauh2019_IJC,tarek_denis1,raissi_at}.

In the case of linear uncertain systems, this cooperativity property simplifies to an elementwise sign condition for the off-diagonal elements of the lower interval bounds of the uncertain dynamics matrix $\mathbf{A} \in \intv{\underline{\mathbf{A}}}{\overline{\mathbf{A}}}$ in the dynamic system model
\begin{equation}\label{eq:sys1}
	\dot{\mathbf{x}}(t) = 
	\mathbf{A} \cdot \mathbf{x}(t)
	\enspace. 
\end{equation}

Guaranteed bounds for the state trajectories $\mathbf{x}(t)$ can be obtained in the form $\sqb{\mathbf{x}}(t) = \intv{\mathbf{v}(t)}{\mathbf{w}(t)}$ so that all reachable states are included by an evaluation of the following coupled set of state equations (which are a direct consequence of~\cite{mueller:1927})
\begin{equation}\label{eq:bracket1}
	\begin{split}
		\dot{v}_i(t) & 
		= 
		\inf\rb{{\sqbs{{A}_{i,i}}} \cdot v_i(t)} + 
		\inf\rb{\sum\limits_{j=1 \atop j\neq i}^n
			{\sqbs{{A}_{i,j}}} \cdot {\sqb{x_j}(t)}} \\
		\dot{w}_i(t) & 
		= 
		\sup\rb{{\sqbs{{A}_{i,i}}} \cdot w_i(t)} + 
		\sup\rb{\sum\limits_{j=1 \atop j\neq i}^n
			{\sqbs{{A}_{i,j}}} \cdot {\sqb{x_j}(t)}}
	\end{split}
\end{equation}
with the resulting state bounds
\begin{equation}
	{\sqb{{x}_i}(t) = \intv{{v}_i(t)}{{w}_i(t)}}\enspace . 
\end{equation}

Note that couplings between the vectors $\mathbf{v}(t)$ and $\mathbf{w}(t)$ disappear in both summations for $j\neq i$ included in~(\ref{eq:bracket1}) if the system is positive, i.e., $v_i(t)\ge 0$ holds for all $t\ge t_0$, see for example \cite{tarek_denis1,tarek_denis2}. There, this property is exploited for the design of a Luenberger-type interval observer, which allows for guaranteed state estimation in the case of bounded measurement uncertainty.

Moreover, recent developments of the so-called mixed monotonicity approach and remainder form decompositions \cite{10057091,9867741} can be exploited as a relaxation of the aforementioned evaluation technique based on cooperativity. In this case, a cooperative subsystem is extracted from the dynamic system model, while the non-cooperative part is {enclosed with the help of} an additive bounded error term. In such a way, the possibility for extracting bounding systems is preserved and the need for computing with general sets is still avoided.

{For scenarios, in which overestimation cannot be avoided fully, the authors of~\cite{Rauh_Auer_2010} suggest the use of systematic criteria to quantify the worst-case level of arising pessimism in the computed state enclosures. Such criteria can be embedded into the development of automatic recommender systems that support users to select the most appropriate set-based solution procedure from a variety of different alternatives~\cite{VERICOMP}.}

\subsection{{Software Libraries for Interval and Set-Based Calculus}}
For {the} practical implementation of set-valued calculus, numerous software libraries exist. The following short list is given without any claim for completeness: \textsc{IntLab} \cite{Rump:1999_1} for \textsc{Matlab} and \textsc{Octave}, the \textsc{Octave} \texttt{interval} package \cite{octave}, 
\textsc{Versoft} \cite{versoft},
\textsc{Juliaintervals} \cite{JuliaIntervals},
\textsc{C-XSC} \cite{Rauh_XSC_URL},
\textsc{DynIbex} \cite{sandretto2016dynbex},
\textsc{Ariadne} \cite{collins-adhs2012},
the interval arithmetic package in \textsc{Mathematica} \cite{Keiper1993IntervalAI},
the \textsc{CORA} library \cite{ARCH15:An_Introduction_to_CORA}, and
\textsc{codac} \cite{codac}. Many of these libraries provide a large number of basic operations, higher level functionalities such as solving linear or nonlinear algebraic sets of equations as well as (global) optimization problems, solving initial value problems for ordinary differential equations, or procedures for the verification of hybrid dynamic models arising in the field of cyber-physical systems. Note, specific approaches that are equally applicable for offline and online use for control design and implementation according to the following subsection are mostly left to the interested users who have to implement these procedures by themselves.

\subsection{{Distinction Between the Offline and Online Use of Set-Based Methods}}
{As far as control engineering applications are concerned, set-valued calculus is generally applicable during two different phases. On the one hand, an \emph{offline} use is possible. In this frame, \emph{offline} means the use during the design phase to verify certain system properties that should be respected by, for example, a control law or state estimation procedure to be developed. On the other hand, the \emph{online} use aims at exploiting set-valued calculus at run-time of a dynamic system, typically coming along with constraints for real-time capability. Then, possible tasks are real-time capable gain scheduling, state estimation, or fault detection.}

\subsection{{Outline of this Article}}
Sec.~\ref{sec:ident} gives an overview of the fundamentals of identifying parameters for dynamic systems with the help of interval and other set-based approaches. These identification results are the basis for the variable-structure control approaches reviewed in Sec.~\ref{sec:control_var} as well as for the guaranteed model predictive control approaches in Sec.~\ref{sec:mpc}. Thereafter, we present approaches for set-based state estimation in Sec.~\ref{sec:IntervalObserverLR}, applicable to the identification of the open-circuit voltage characteristic of Lithium-ion batteries, and an ellipsoidal technique in Sec.~\ref{sec:ellipse} that allows for treating set-valued and stochastic uncertainty and disturbances in a unified manner when solving state estimation tasks. Finally, conclusions and a brief outlook on future work are given in Sec.~\ref{sec:concl}.

\section{System Modeling and Verified Parameter Identification}\label{sec:ident}
The estimation of the parameters of both finite-dimensional sets of discrete-time difference equations as well as continuous-time ordinary differential equations is crucial for the design and implementation of all model-based control and state estimation procedures. Identification experiments are typically performed (for asymptotically stable systems) with predefined, time-dependent actuation signals so that measurements of selected outputs can be gathered at certain discrete points of time $t_k$, $k \in \mathbb{N}_0$. Here, $t_0=0$ is assumed without loss of generality. Due to the collection of measurements at discrete points of time, it is necessary that parameter identification routines allow for a propagation of state information between two subsequent measurement instants $t_{k-1}$ and $t_k$, $k \ge 1$. The following description of the set-based parameter identification approach is based on the more detailed discussions that can be found in \cite{RAUH2020112484}.

\subsection{Predictor--Corrector Framework for the Verified Identification of Parameters of Dynamic Systems}

\begin{figure}[htp]
	\centering
	{\resizebox{0.8\linewidth}{!}{
			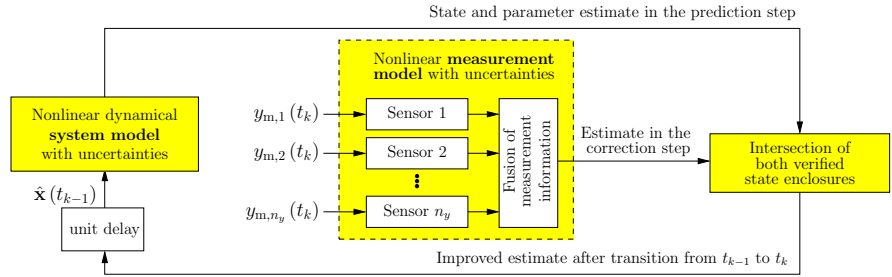}}
	\caption{\label{fig:appr1}Predictor--corrector scheme for combined state estimation and parameter identification \cite{RAUH2020112484}.}
\end{figure}

As summarized in Fig.~\ref{fig:appr1}, a first class of verified parameter identification routines is characterized by an observer-based structure. This structure is closely linked to the evaluation steps that are performed in classical state estimators such as Luenberger-type observers or (Extended) Kalman Filters~\cite{Kalman:1960,Julier:2000}. Hence, prediction and correction steps are separated from each other and evaluated in an alternating manner.

\begin{figure}[htp]
	\centering
	\subfloat[Measurement information of the system output $y_\mathrm{m}(t)$, sampled at the equidistant instants $t_0, t_1, t_2, \ldots$.  ]{\includegraphics[width=0.485\linewidth]{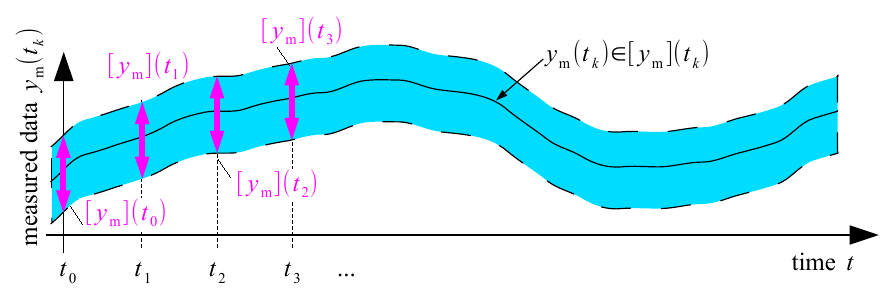}}\hfill
	\subfloat[Output trajectory including interval uncertainty for a possibly feasible (undecided) system parameterization (\textbf{Case~1}). ]{\includegraphics[width=0.485\linewidth]{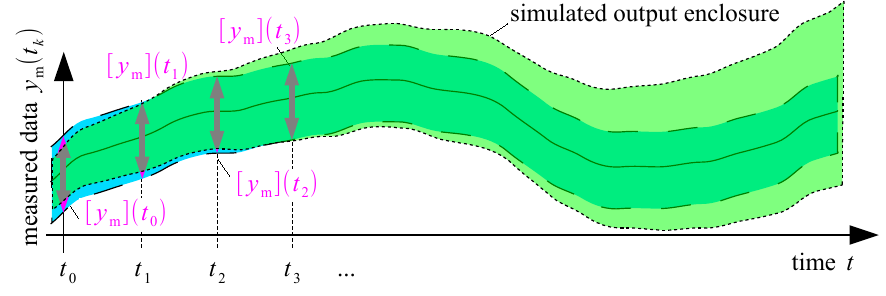}}\\
	\subfloat[Output trajectory including interval uncertainty for a guaranteed feasible system parameterization (\textbf{Case~2}).]{\includegraphics[width=0.485\linewidth]{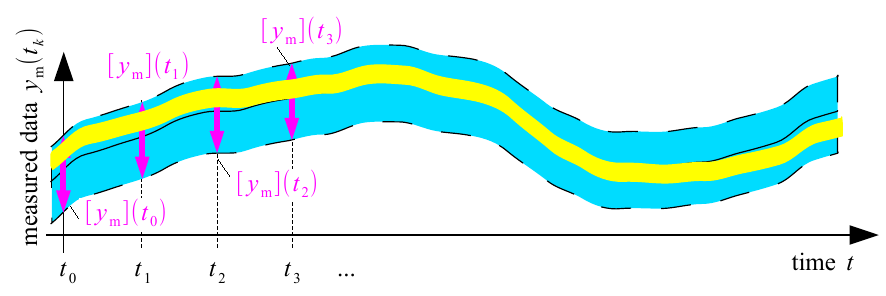}}\hfill
	\subfloat[Output trajectory including interval uncertainty for a guaranteed infeasible system parameterization (\textbf{Case~3}).]{\includegraphics[width=0.485\linewidth]{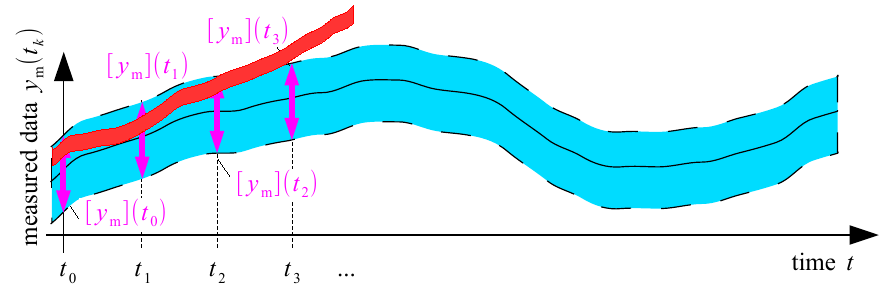}}
	\caption{\label{fig:appr2}Simulation-based distinction between possibly feasible, guaranteed feasible, and infeasible system parameterizations~\cite{rauhdoetschelsofc4,RAUH2020112484}.}
\end{figure}

In this framework, the detection of feasible parameter domains is performed by means of the following three-stage procedure~\cite{rauhdoetschelsofc1,rauhdoetschelsofc4}: 
\begin{enumerate}
	\item A verified evaluation of the set of state equations is performed between two subsequent measurement points. In the case of discrete-time state equations, the corresponding expressions are evaluated with the help of interval arithmetic (or one of the alternative set-valued approaches mentioned in the introduction of this article), while verified solvers for initial value problems for ordinary differential equations \cite{Sandretto:2016:VEI,vspode,Kapela,Ned:2011,AR_2006,Nedialkov_phd,sandretto2016dynbex,Ned:2006,kuehn} are required if continuous-time processes are taken into consideration. In any case, problem-specific approaches for the reduction of overestimation~\cite{Rauh_Krasnochtanova_MMAR2011} --- such as {the exploitation of cooperativity,} interval subdivision, preconditioning of state equations, symbolic simplifications, or enclosures allowing to trace the shape of the reachable state domains in a tight manner --- shall be employed to minimize the pessimism in the computed state and output boundaries that is caused by both the dependency and wrapping effects~\cite{AppliedIntervalAnalysis,Lohner:1987}.
\item Information resulting from the prediction of state intervals up to the point at which new measurements are available are combined with information obtained from sensors. The applicable techniques are:
\begin{itemize}
	\item {An intersection} of the enclosures of predicted state variables with {interval bounds for} directly measured state intervals (in terms of an additive superposition of a point-valued measurement with the respective tolerance interval);
	\item The use of contractors and set inversion approaches to determine an inverse mapping from the measured outputs to the internal system states and parameters \cite{ifqir2022zonotopic}. {Examples} are forward--backward contractors~\cite{AppliedIntervalAnalysis} initialized with the predicted state intervals that are intersected with the associated measurement outputs or interval Newton techniques in the case of nonlinear output equations or for output equations involving more than one state variable. {Additionally, also observer-based formulations for solving the same task were recently published in~\cite{ifqir2022zonotopic}, where zonotopes are employed as the underlying uncertainty representation.}
\end{itemize}
\item Parameter subdomains are eliminated {either by using contractors} or after a subdivision of parameter intervals with associated multiple evaluations of the state equations. In the latter case, a parameter subinterval is guaranteed to be inconsistent with the information provided by the specified system model and the available measurements if the result of the state prediction substituted into the measurement model and the interval-valued measured data do not overlap. 
For a visualization of this exclusion text, see Fig.~\ref{fig:appr2}, where a branch-and-bound procedure on the basis of multiple simulations of the state equations over the complete horizon of gathered {measured data is depicted.} 
\end{enumerate}

\subsection{Simulation-Based Parameter Identification Procedures}

Simulation-based parameter identification procedures, making use of multiple evaluations of the set of state equations for subintervals of the complete initial domain of possible system parameters, are {one of the available options} for parameter identification.

The underlying branch-and-bound procedure\footnote{Branching of the parameter domain by an interval subdivision approach with a subsequent simulation-based bounding of the output trajectories.} relies on the following assumptions:
\begin{itemize}
	\item measured data are available at discrete points of time, 
	\item worst-case bounds for measurement tolerances are known a-priori (modeling of the sensor uncertainty), 
	\item the model structure to be parameterized is assumed to be structurally correct,
	\item outer enclosures for the domains of possibly uncertain initial states are given, and 
	\item interval bounds on the uncertain parameters are known in terms of conservative overapproximations.
\end{itemize}

To exclude parts of the domains of uncertain initial states and parts of the a-priori given parameter intervals, a subdivision procedure of the respective domains is executed~\cite{CSENDES:1997}. To detect intervals for the parameterization of the system model which are guaranteed to be inconsistent, directly measured and simulated state intervals are intersected at identical points of time. If state variables are not directly measured, the contractor procedures mentioned in the previous subsection are equally applicable here. 
Parameter interval subdivisions are performed according to the following distinction of cases, where the directions {in which the parameter intervals are subdivided are determined} with the help of a sensitivity analysis \cite{rauhdoetschelsofc4}. For accelerations by both additional (physics-inspired) contractors and GPU-based parallelizations, the readers are referred to \cite{RAUH2020112484,AUER2020112657}:
\begin{description}
	\item[\textbf{Case~1}] Parameter subintervals are yet undecided if their corresponding predicted output intervals overlap at least partially for all available sensors at each point of time at which measurements are available and are never outside the range of measured data.
	\item[\textbf{Case~2}] A parameter subinterval is guaranteed to be consistent if the corresponding output intervals are true subintervals of the interval-valued measured data for each of the available sensors at each point of time. This subinterval is no longer investigated {in the exclusion strategy} but stored in a list of guaranteed admissible boxes.
	\item[\textbf{Case~3}] 
	A parameter subinterval is guaranteed to be inconsistent if the corresponding output interval lies outside the range of the interval-valued measured data. This has to hold true for at least one of the available sensors and for at least one point in time. The resulting parameter subinterval is excluded from further evaluations and the underlying simulation can be aborted {at that point in time as which the} inconsistency is {firstly} detected.
\end{description}

 \section{{Interval-Based Variable-Structure Control}}\label{sec:control_var}
 As mentioned in the previous section, verified parameter identification is the prerequisite for the implementation of guaranteed stabilizing control procedures. These include linear matrix inequality methods for systems with polytopic and norm-bounded parameter uncertainty \cite{LMIs,Rauh2019_IJC}, interval-based gain scheduling techniques \cite{Kersten2018:GC}, and variable structure as well as backstepping-type controllers \cite{Rauh_Senkel_springer_2015,Rauh_Senkel_springer_2015b,Rauh_Rauh_NOLCOS:2013_IMA,Rauh_Rauh:2015_TIE} without and with hard state constraints.

In this section, we review a robust control procedure that extends the sliding mode design methodology by means of the online use of interval analysis to prevent the violation of state constraints with certainty \cite{Rauh_Senkel_springer_2015,Rauh_Senkel_springer_2015b,7739793}. For a similar approach, applicable to the backstepping design methodology, the reader is referred to \cite{RAUH20176283}.

 \subsection{{Goal of the Control Synthesis}}{Variable-structure control procedures aim at solving two-fold control tasks. First, they ensure that a system state is guided towards a time-dependent desired trajectory within finite time. Second, they guarantee that this desired trajectory is tracked with high accuracy as soon as it has been reached. A possible application scenario is heating up a solid oxide fuel cell stack from ambient temperature to its usual operating point with temperatures above $800\,\mathrm{K}$ along a pre-defined trajectory and keeping the temperature in close vicinity of that operating point despite variations of the electric load taken from the fuel cell. Accurate temperature control, including a guaranteed prevention of a rise of the temperature over a maximum admissible value is crucial to simultaneously ensure a high conversion efficiency of the chemical energy of hydrogen used as the fuel gas and simultaneously avoiding accelerated system degradation and hazardous states caused by overtemperatures~\cite{Rauh_Senkel_springer_2015b,Rauh_Rauh:2015_TIE}.}

 \subsection{Fundamental Sliding Mode Control Laws}
  
 Both the treatment of strict inequality constraints {representing the limitations of the admissible operating conditions such as the temperature limit above} and bounded interval uncertainty {of internal system parameters} can be combined with {classical} first- and second-order sliding mode techniques \cite{Rauh_Eker2010,Rauh_Fridman}. These classical approaches have the drawback that they suppose perfectly known parameters and can therefore typically not ensure the guaranteed satisfaction of state constraints. Moreover, they may be inefficient in the use of the available ranges of actuator signals~\cite{6347818}.
  
 \subsubsection{First-Order Sliding Mode Control}
For a summary of the fundamental sliding mode control design of single-input single-output systems, consider the $n$-th order set of ordinary differential equations to be given in the nonlinear controller canonical form
  \begin{equation}\label{Rauh_eq:ex_state}
  	\begin{split}
  		\dot{\mathbf{x}}(t) & \! = \!
  		\begin{bmatrix}
  			\dot{x}_1(t) &
  			\ldots &
  			\dot{x}_{n-1}(t) &
  			\dot{x}_n(t)
  		\end{bmatrix}^T \\
  		& \! = \!
  		\begin{bmatrix}
  			{x}_2(t) &
  			\ldots &
  			{x}_{n}(t) &
  			a\rb{{\mathbf{x}}(t),{\mathbf{p}}} + b\rb{{\mathbf{x}}(t),{\mathbf{p}}} \! \cdot \! v(t)
  		\end{bmatrix}^T
  	\end{split}
  	\raisetag{11mm}
  \end{equation}
with the state vector ${\mathbf{x}}(t) \in \mathbb{R}^n$ and the control input {$v(t)\in \mathbb{R}$. To ensure controllability, we assume $b\rb{{\mathbf{x}}(t),{\mathbf{p}}} \neq 0$. Under this assumption, a feedback linearizing control strategy}
  \begin{equation}\label{eq:lin_contr}
  	v(t) = \frac{-a\rb{{\mathbf{x}}(t),{\mathbf{p}}} + u(t)}{b\rb{{\mathbf{x}}(t),{\mathbf{p}}}}
  \end{equation}
  {can be designed. It aims at compensating the nonlinear expressions $a\rb{{\mathbf{x}}(t),{\mathbf{p}}}$ and $b\rb{{\mathbf{x}}(t),{\mathbf{p}}}$ which both depend on the state ${\mathbf{x}}(t)$ and internal parameters ${\mathbf{p}}$. However, in reality, these functions are imperfectly known so that the auxiliary input $u(t)$ needs to ensure both robustness and stability despite uncertainty.}
  
  For this system model, the output variable {to be controlled} is given by 
  \begin{equation}\label{Rauh_eq:ex_output}
  	y(t) = x_1(t)
  	\enspace .
  \end{equation}
  
  Due to the assumed structure of the system~(\ref{Rauh_eq:ex_state}), the output variable~(\ref{Rauh_eq:ex_output}) has the relative degree $n$~\cite{Rauh_Marquez:2003}, corresponding to the property that the $n$-th time derivative $x_1^{(n)}(t)$ of the system output is the lowest-order derivative explicitly depending on either of the control inputs $v(t)$ and $u(t)$. Therefore, the output $y(t)$ corresponds to a (trivial) flat system\footnote{A system is differentially flat if both the control signal and the state trajectories can be expressed purely in terms of a flat output (not necessarily coinciding with the output to be controlled) and a finite number of time derivatives thereof.} output~\cite{Rauh_Fliess}. {This property of differential flatness ensures that} an exact tracking of a sufficiently smooth{, i.e., $n$ times continuously differentiable,} desired trajectory~$x_{1,\mathrm{d}}(t)$ can be achieved in the case of exactly known parameters. 
  
  On the basis of this desired trajectory, the corresponding tracking error 
    \begin{equation}
  	{\xi}_1(t)
  	= x_1(t) - x_{1,\mathrm{d}}(t)
  \end{equation}
  and its $r$-th time derivative\footnote{Time arguments are omitted {subsequently}, whenever the meaning is non-ambiguous.} are {introduced. To represent both in a compact form, we exploit the notation} 
  \begin{equation}
  {\xi}_1^{(r)}(t)
  = x_1^{(r)}(t) - x_{1,\mathrm{d}}^{(r)}(t)~,~~r \in \{0,1,\ldots, n\}\enspace ,
  \end{equation}
  {to define} the sliding surface\footnote{{In this context, sliding means that the system dynamics obey a differential equation of order smaller than $n$ as soon as the state is bound to the surface $s=0$, cf.~\cite{Utkin1992SlidingMI}.}}
  \begin{equation}
 { s(t) = \sum\limits_{r=0}^{n-1}
  \alpha_r {\xi}_1^{(r)}(t)}
  \end{equation}
  in which the highest-order coefficient is normalized according to $\alpha_{n-1}=1$. {Note that in the case of perfect trajectory tracking, the equality $s(t)=0$ holds.}
  
  To guarantee asymptotic stability for states exactly located on this sliding surface, the parameters $\alpha_r$ have to fulfill the necessary and sufficient stability conditions for a Hurwitz polynomial of the order $n-1$. 
  
  A classical first-order sliding mode control can be derived with the help of the quadratic radially unbounded candidate for a {positive definite} Lyapunov function\footnote{{The superscript ${\dbr{\mathrm{I}}}$ for the Lyapunov function and the associated control law serve as a distinction from the second-order procedure discussed in the following subsection.}}
  \begin{equation}
  	{V^{\dbr{\mathrm{I}}}(t) = \frac{1}{2} s^2(t)}
  	\enspace ,
  \end{equation}
  where $s(t)$ depends on the tracking error and its derivatives and therefore is a function of the system state, see also the definition (\ref{Rauh_eq:ex_state}).
  
  {For $s \neq 0$, $V^{\dbr{\mathrm{I}}} >0$ holds. State trajectories leading to the desired system behavior $V=0$ can be obtained if (global) asymptotic stability of the dynamic system can be proven. This stability requirement }corresponds to the (global) negative definiteness of the corresponding time derivative
  \begin{equation}\label{Rauh_eq:ex_dV}
  	\dot{V}^{\dbr{\mathrm{I}}}{(t)} = {s(t) \cdot \dot{s}(t) = 
  	\rb{\sum\limits_{r=0}^{n-1}
  		\alpha_r {\xi}_1^{(r)}(t)} \cdot
  	\rb{\sum\limits_{r=0}^{n-1}
  		\alpha_r {\xi}_1^{(r+1)}(t)}} < 0
  	\enspace ,
  \end{equation}
  {where $\dot{s}$ explicitly depends on the control input.}
  
  {The stability requirement above ensures that the system state converges towards the desired reference if $s \neq 0$, either due to initial conditions not perfectly coinciding with the reference or due to the occurrence of disturbances.}
  
  Following the derivation of the variable-structure control law according to~\cite{Rauh_Rauh_NOLCOS:2013_IMA,Rauh_Rauh:2015_TIE}, the right-hand side of the inequality~(\ref{Rauh_eq:ex_dV}) is replaced by the more conservative formulation
  \begin{equation}\label{Rauh_eq:ex_dV1}
  		\rb{\sum\limits_{r=0}^{n-1}
  			\alpha_r {\xi}_1^{(r)}} \cdot
  		\rb{\sum\limits_{r=0}^{n-1}
  			\alpha_r {\xi}_1^{(r+1)}}< - \eta \cdot \left|s\right| = 
  		-\eta \cdot \rb{\sum\limits_{r=0}^{n-1}
  			\alpha_r {\xi}_1^{(r)}}
  		\cdot \mathrm{sign}\rb{s}
  \end{equation}
  which guarantees global asymptotic stability for $\eta>0$. Here, the actual choice of {the constant} $\eta$ significantly influences the dynamics and the maximum absolute values of the control signal in the reaching phase characterized by $s \neq 0$. As soon as $s=0$ has been reached after a finite time duration, the control amplitudes depend (in the unperturbed case) on the actual choice of {the reference trajectory $x_{1,\mathrm{d}}$.}
  
   {The time-independent coefficients $\alpha_r$ have a major influence} on the control amplitudes if non-modeled errors and disturbances act on the system dynamics and if the error signals ${\xi}_1^{(r)}$ are corrupted by non-negligible measurement noise or state reconstruction errors, preventing the perfect tracking of the desired trajectory~$x_{1,\mathrm{d}}(t)$ and making it impossible that $s\equiv 0$ holds perfectly after the end of the reaching phase. 
  
  The derivation of the control law is completed by enforcing that the second factor in~(\ref{Rauh_eq:ex_dV1}) becomes proportional to the sign of the actual value of $s$ according to
  \begin{equation}\label{Rauh_eq:u_I_tmp}
  			\sum\limits_{r=0}^{n-2}
  			\alpha_r {\xi}_1^{(r+1)}
  			+ u - x_{1,\mathrm{d}}^{(n)} + \eta \cdot \mathrm{sign}\rb{s}
  			=-\beta \cdot \mathrm{sign}\rb{s}
  			%
  		\quad \text{with {an arbitrary}} \quad
  		\beta>0\enspace .
  \end{equation}
  Using \mbox{$\tilde{\eta}:=\eta+\beta>0$}, the control law \mbox{results in}
  \begin{equation}\label{Rauh_eq:u_I_SM}
  	{u} =
  	u^{\dbr{\mathrm{I}}} = 
  	x_{1,\mathrm{d}}^{(n)}
  	-\sum\limits_{r=0}^{n-2}
  	\alpha_r {\xi}_1^{(r+1)}
  	- \tilde{\eta} \cdot \mathrm{sign}\rb{s}
  	\enspace. 
  \end{equation}
  
  \subsubsection{Second-Order Sliding Mode Control}
  
  For a second-order sliding mode, both $s=0$ and $\dot{s}=0$ have to be ensured by the feedback controller~\cite{Rauh_Eker2010,Rauh_Fridman}. This can be achieved by adding a first-order lag dynamics on the left-hand side of
  \begin{equation}\label{Rauh_eq:ex_s_second_order2}
  	\gamma_1 \dot{s} + \gamma_0 s =
  	\sum\limits_{r=-1}^{n-1}
  	\alpha_r {\xi}_1^{(r)}
  	\quad  \text{with} \quad 
  	{\xi}_1^{(-1)}(t) := \int_0^t{\xi}_1(\tau)\mathrm{d}\tau
  	\, .
  \end{equation}
  For the sake of asymptotic stability, the coefficients $\gamma_0$ and $\gamma_1$ need to be strictly positive. For $\alpha_{-1}=0$, this sliding surface has a proportional and differentiating characteristic, while $\alpha_{-1}\neq 0$ means that the integral ${\xi}_1^{(-1)}$ of the output error is additionally fed back in the definition of the sliding variable $s$.
  
  To ensure both $s=0$ and $\dot{s}=0$ for the closed-loop control system, the Lyapunov function candidate, which was chosen {in the previous subsection} as a function solely depending on the value of $s$, is redefined in the form
  \begin{equation}\label{eq:lyap_II}
  	V^{\dbr{\mathrm{II}}} = \frac{1}{2} \cdot \rb{
  		s^2 + \lambda \dot{s}^2}
  	\quad \text{with the scaling factor} \quad
  	\lambda > 0
  	\enspace .
  \end{equation}
  {This extended specification typically enhances the closed-loop system's robustness by the introduction of the temporal integral of the error signal according to~(\ref{Rauh_eq:ex_s_second_order2}) in the definition of the sliding surface. In practical applications, $\lambda$ is chosen as shown subsequently so that the stability requirements are simplified.}
  
  The computation of the time derivative of~(\ref{eq:lyap_II}) results in 
  \begin{equation}\label{Rauh_eq:dV_ext}
  	\dot{V}^{\dbr{\mathrm{II}}} = s \cdot  \dot{s} + \dot{s} \cdot \rb{
  		- \frac{\lambda \gamma_0}{\gamma_1} \dot{s} 
  		+ \frac{\lambda}{\gamma_1}
  		\sum\limits_{r=0}^{n}
  		\alpha_{r-1} {\xi}_1^{(r)}
  	} < 0
  	\enspace ,
  \end{equation}
  where {the specific parameter choice} $\lambda = \gamma_1 >0$ is used {in the following} without loss of generality~\cite{Rauh_Senkel_springer_2015b}. Also according to~\cite{Rauh_Senkel_springer_2015b}, the stabilization of the closed-loop system towards $s=0$ can be achieved by {requiring} 
  \begin{equation}
  	\dot{V}^{\dbr{\mathrm{II}}} < - \eta_1\cdot \left|\dot{s}\right|
  	- \eta_2\cdot \left|{s}\right|\cdot \left|\dot{s}\right|
  	= - \dot{s} \cdot \mathrm{sign}\rb{\dot{s}} \cdot \rb{
  		\eta_1 + \eta_2\cdot \left|{s}\right|
  	}
  	.
  \end{equation}
  This finally leads to the nonlinear feedback controller~\cite[generalized form of Eqs.~(22), (23)]{Rauh_Eker2010}
  \begin{equation}\label{eq:second_order_sm}
  	u = 
  		u^{\dbr{\mathrm{II}}} = x_{1,\mathrm{d}}^{(n)} +
  		\frac{1}{\alpha_{n-1}}
  		\cdot \Biggl(\gamma_0 \dot{s} - s 
  		- \sum\limits_{r=0}^{n-1}
  		\alpha_{r-1} {\xi}_1^{(r)}
  		  		- \mathrm{sign}\rb{\dot{s}} \cdot \rb{
  			\tilde{\eta}_1 + \tilde{\eta}_2\cdot \left|{s}\right|
  		}\Biggr)
  \end{equation}
  with $\tilde{\eta}_i \ge \eta_i >0$ for both $i \in \{1,2\}$. {In analogy to the previous subsection, constant design parameters are the gains $\tilde{\eta}_i$ of both variable-structure control components as well as the coefficients $\alpha_{r-1}$ together with $\gamma_0$.}

  \subsection{Extension by One- and Two-Sided Barrier Functions}
  Both control laws $u^{\dbr{\mathrm{I}}}$ {(first-order sliding mode)} and $u^{\dbr{\mathrm{II}}}$ {(second-order sliding mode)} can be extended by one- and two-sided barrier functions{, which are subsequently distinguished by the superscripts ${\dbr{\mathrm{A}}}$ and ${\dbr{\mathrm{B}}}$}~\cite{Rauh_barrier}.
  
 \subsubsection{One-Sided State Constraints} 
  For the case of a one-sided barrier {that represents an upper bound for the output variable}, it is necessary that the constraint\footnote{{For cases, where constraints on other interval state variables are specified, the reader is referred to \cite{Rauh_barrier}.}}
  \begin{equation}\label{Rauh_eq:ex_constraint}
  	{x_1(t) < \bar{x}_{1,\mathrm{max}}(t) := x_{1,\mathrm{d}}(t) + \Delta{x}_{1,\mathrm{max}}}
  	\quad\text{with} \quad
  	\Delta{x}_{1,\mathrm{max}} >0
  \end{equation}
  is not violated for any point of time $t>0$. {As a prerequisite, we assume that} the initial conditions for the state vector ${\mathbf{x}}(t)$ at $t=0$ are compatible with this constraint {as well}. Moreover, it is necessary that the sliding surface $s=0$ (equivalent to $x_{1}=x_{1,\mathrm{d}}$) lies within the admissible operating range defined in~(\ref{Rauh_eq:ex_constraint}). {Here, $\bar{x}_{1,\mathrm{max}}$ denotes the maximum admissible upper bound for the state $x_1$, while $\Delta{x}_{1,\mathrm{max}}$ is a safety margin expressing the minimum difference between the reference signal and the threshold value.}
  
  Then, the extended Lyapunov function ansatz
  \begin{equation}\label{Rauh_eq:V_A}
  	\begin{split}
  		{V}^{\dbr{j,\mathrm{A}}} & = V^{\dbr{j}} + V^{\dbr{\mathrm{A}}}
  		>0
  		\quad \text{for} \quad
  		s \neq 0
  		\quad \text{with}
  	\end{split}
  \end{equation}
  \begin{equation}\label{Rauh_eq:ex_barrier}
  	V^{\dbr{\mathrm{A}}} = 
  	\rho_\mathrm{V} \cdot
  	\ln\rb{\tfrac{\sigma_\mathrm{V} \cdot \bar{x}_{1,\mathrm{max}}}{\bar{x}_{1,\mathrm{max}}-x_1}} 
  	~ \text{and} ~
  	x_1 < \bar{x}_{1,\mathrm{max}}
  \end{equation}
  is introduced for both alternatives $j \in \{\mathrm{I},\mathrm{II}\}$. In~(\ref{Rauh_eq:ex_barrier}), $\rho_\mathrm{V}>0$ is selected so that the singularity ${\bar{x}_{1,\mathrm{max}}-x_1}=0$ represents a repelling potential and that control constraints are not violated for usual operating conditions. Then, the term $V^{\dbr{j}}$ has dominating influence in the neighborhood of~$s=0$, while $\sigma_\mathrm{V}>0$ can be used to adapt the steepness of the barrier function near its singularity.
  
  In the case of the first-order sliding mode, the time derivative of~(\ref{Rauh_eq:V_A}) can be computed as
  \begin{equation}\label{Rauh_eq:ex_dV_extended1}
  		\dot{V}^{\dbr{j,\mathrm{A}}} = \dot{V}^{\dbr{j}} + \dot{V}^{\dbr{\mathrm{A}}}
  		<0 \quad \text{with} \quad
  		\dot{V}^{\dbr{\mathrm{A}}} = \tfrac{\rho_\mathrm{V}}{\bar{x}_{1,\mathrm{max}}} \cdot \rb{\tfrac{-x_1  \cdot 
  				\dot{\bar{x}}_{1,\mathrm{max}} + \dot{x}_1  \cdot \bar{x}_{1,\mathrm{max}}}{\bar{x}_{1,\mathrm{max}}-x_1}}
  		\enspace ,
  \end{equation}
  where $\dot{V}^{\dbr{\mathrm{A}}}$ does not explicitly depend on the system input $u$.
  
  In analogy to the fundamental first-order sliding mode control law $u^{\dbr{\mathrm{I}}}(t)$ derived from~(\ref{Rauh_eq:u_I_tmp}), the inequality 
  \begin{equation}\label{Rauh_eq:ex_dV_extended2}
  	s
  	\cdot
  	\underbrace{
  		\rb{\sum\limits_{r=0}^{n-2}
  			\alpha_r {\xi}_1^{(r+1)}
  			+ u - x_{1,\mathrm{d}}^{(n)} + \eta \cdot \mathrm{sign}\rb{s}
  			+ \frac{1}{s} \cdot \dot{V}^{\dbr{\mathrm{A}}}
  	}}_{-\beta \cdot \mathrm{sign}\rb{s}} < 0
  \end{equation}
  {needs to be satisfied}. Following the same steps as in the derivation of $u^{\dbr{\mathrm{I}}}(t)$ yields the control law
  \begin{equation}\label{Rauh_eq:ex_dV_extended3}
  	u={u}^{\dbr{\mathrm{I},\mathrm{A}}} = u^{\dbr{\mathrm{I}}} - \tfrac{s}{s^2 + \tilde{\epsilon}} \cdot  \dot{V}^{\dbr{\mathrm{A}}}
  	\enspace,
  \end{equation}
  in which $\frac{1}{s}$ in~(\ref{Rauh_eq:ex_dV_extended2}) has been approximated by $\frac{s}{s^2 + \tilde{\epsilon}}$ with the small constant $\tilde{\epsilon}>0$ to ensure regularity of the control law ${u}^{\dbr{\mathrm{I},\mathrm{A}}}(t)$ on the sliding surface $s=0$. Moreover, this modification guarantees that the barrier function becomes inactive as soon as the control goal has been reached. 
  
  In analogy to~(\ref{Rauh_eq:ex_dV_extended3}), {also} the second-order sliding mode control~(\ref{eq:second_order_sm}) can be extended by the barrier function~(\ref{Rauh_eq:ex_barrier}). Following the same steps as in Eqs.~(\ref{Rauh_eq:ex_dV_extended1})--(\ref{Rauh_eq:ex_dV_extended3}) leads to
  \begin{equation}\label{eq:u_II}
  	u={u}^{\dbr{\mathrm{II},\mathrm{A}}} = u^{\dbr{\mathrm{II}}} - 
  	\tfrac{1}{\alpha_{n-1}} \cdot 
  	\tfrac{\dot{s}}{\dot{s}^2 + \tilde{\epsilon}} \cdot  \dot{V}^{\dbr{\mathrm{A}}}
  	\enspace .
  \end{equation}

 \subsubsection{Two-Sided State Constraints} 
  For the case of two-sided state constraints, the Lyapunov functions $V^{\dbr{j}}$, $j \in \{\mathrm{I},\mathrm{II}\}$, are extended according to
  \begin{equation}\label{Rauh_eq:ex_Lyap_2side}
  	{V}^{\dbr{j,\mathrm{B}}} = V^{\dbr{j}} + V^{\dbr{\mathrm{B}}}
  	>0
  	\quad \text{for} \quad
  	s \neq 0
  	\enspace ,
  \end{equation}
 where the additive term $V^{\dbr{\mathrm{B}}}$ is chosen in this paper so that state deviations {$\left|x_1(t)-x_{1,\mathrm{d}}(t)\right| \ge \bar{\chi}$} are {guaranteed to be avoided~\cite{Rauh_Senkel_springer_2015b}.} Symmetric barriers can be enforced by specifying a barrier function in the form
  \begin{equation}\label{Rauh_eq:ex_barrier_2side}
  	V^{\dbr{\mathrm{B}}} = 
  	\rho_\mathrm{V} \cdot  \ln\rb{\tfrac{\bar{\chi}^{2l}}{\bar{\chi}^{2l}-\rb{x_1-x_{1,\mathrm{d}}}^{2l}}}
  	\quad \text{with} \quad
  	l \in \mathbb{N}
  	\enspace .
  \end{equation}
  Large values for $l$ typically lead to the fact that resulting state trajectories come closer to the edges of the admissible operating range. As before, the time derivative of the additive term~(\ref{Rauh_eq:ex_barrier_2side}) is computed, which yields 
  \begin{equation}
  	\dot{V}^{\dbr{\mathrm{B}}} = 
  	\rho_\mathrm{V} \cdot \tfrac{{2l} \cdot \rb{x_1-x_{1,\mathrm{d}}}^{2l-1} \cdot \rb{\dot{x}_1-\dot{x}_{1,\mathrm{d}}}}{\bar{\chi}^{2l}-\rb{x_1-x_{1,\mathrm{d}}}^{2l}}
  	\enspace .
  \end{equation}
In full analogy to Eqs.~(\ref{Rauh_eq:ex_dV_extended1})--(\ref{eq:u_II}), the requirement $\dot{V}^{\dbr{j,\mathrm{B}}}<0$ for $s \neq 0$ (and $\dot{s}\neq 0$, resp.) leads to the extension 
  \begin{equation}\label{Rauh_eq:u_I_B}
  	{u}^{\dbr{\mathrm{I},\mathrm{B}}} = u^{\dbr{\mathrm{I}}} - s^{-1} \cdot \dot{V}^{\dbr{\mathrm{B}}} 
  \end{equation}
  of the first-order sliding mode controller and to
  \begin{equation}
  	{u}^{\dbr{\mathrm{II},\mathrm{B}}} = u^{\dbr{\mathrm{II}}} - \rb{\alpha_{n-1} \cdot \dot{s}}^{-1} \cdot \dot{V}^{\dbr{\mathrm{B}}}
  \end{equation}
  for the second-order case, where the same regularization strategies for the rational terms $\frac{1}{s}$ and $\frac{1}{\dot{s}}$ become necessary as before. {For the implementation of the control law, the} term $\dot{s}$ (if required) is typically estimated {at run-time} by a low-pass filtered differentiation or by means of an observer.

  \subsection{Interval Extensions to Handle Bounded Parameter Uncertainty and State Estimation Errors}\label{Rauh_sec:ISMC}
  
  To guarantee asymptotic stability despite bounded uncertainty in both the measured (resp., estimated) states {$\mathbf{x}(t)$} and system parameters~${\mathbf{p}}$ included in the control law, interval techniques can be applied at run-time during the execution of the previously derived control laws if they are extended according to~\cite{Rauh_Rauh:2015_TIE,Rauh_Senkel_springer_2015b}. 
  {We assume in the following that a suitable interval observer (e.g., \cite{MazBern,raissi_at}) provides the interval bounds}  ${\mathbf{x}} \in \sqb{{\mathbf{x}}} = \intv{\ul{\mathbf{x}}}{\ol{\mathbf{x}}}$ {for the state vector or that all components of this vector are directly measured with bounded uncertainty}. In addition, it is assumed that the system model is given as an $n$-th order set of ODEs (\ref{Rauh_eq:ex_state}) 
  in nonlinear controller canonical form. If this is not the case, extended control approaches according to~\cite{Rauh_Senkel_springer_2015} are applicable.

  For the sake of controllability (and, hence, also for the existence of the following interval-based variable-structure controllers), it has to be guaranteed that
  \begin{equation}\label{eq:ctrb}
  	{0 \not\in \sqb{b}\rb{\sqb{\mathbf{x}},\sqb{\mathbf{p}}}}{, \quad \forall {{\mathbf{x}} \in \sqb{{\mathbf{x}}}} \quad \text{and} \quad \forall{\mathbf{p}} \in \sqb{{\mathbf{p}}}}
  \end{equation}
  holds as already discussed for Eq.~(\ref{eq:lin_contr}). The constraint (\ref{eq:ctrb}) is a necessary controllability requirement that needs to be satisfied along all prossible state trajectories $\mathbf{x}(t)$ as well as for all possible system parameters. To handle the set-valued state and parameter uncertainty, the output tracking error and its $r$-th derivative are enclosed by the intervals
  \begin{equation}\label{Rauh_eq:ex_error_int}
  	{{\xi}_1^{(r)}(t) \in \sqb{{\xi}_1^{(r)}}(t)
  	= \sqb{x_1^{(r)}}(t) - x_{1,\mathrm{d}}^{(r)}(t)}
  \end{equation}
  for each $r \in \{0,1,\ldots, n\}$. These tracking error intervals can be used to generalize the first-order variable-structure controller (without and with state constraints) according to
  \begin{equation}\label{Rauh_eq:u_interval1}
  	\sqb{v^{\dbr{\mathrm{I}}}} = 
  	{\tfrac{-\sqb{a}\rb{\sqb{{\mathbf{x}}},\sqb{{\mathbf{p}}}} + 
  		x_{1,\mathrm{d}}^{(n)}
  		-\sum\limits_{r=0}^{n-2}
  		\alpha_r \cdot \sqb{{\xi}_1^{(r+1)}}
  		- \tilde{\eta} \cdot \mathrm{sign}\rb{\sqb{s}}}{\sqb{b}\rb{\sqb{\mathbf{x}},\sqb{\mathbf{p}}}}}
  	\enspace,
  \end{equation}
  \begin{equation}\label{Rauh_eq:u_interval1a}
  	{\sqb{{v}^{\dbr{\mathrm{I},\mathrm{A}}}} = \sqb{v^{\dbr{\mathrm{I}}}} - \tfrac{1}{\sqb{b}\rb{\sqb{\mathbf{x}},\sqb{\mathbf{p}}}} \cdot \tfrac{\sqb{s}}{\sqb{s}^2 + \tilde{\epsilon}} \cdot  \sqb{\dot{V}^{\dbr{\mathrm{A}}}}}
  	\enspace ,
  \end{equation}
  and
  \begin{equation}\label{Rauh_eq:u_interval1b}
  	{\sqb{{v}^{\dbr{\mathrm{I},\mathrm{B}}}} = 
  	\sqb{v^{\dbr{\mathrm{I}}}} - \tfrac{1}{\sqb{b}\rb{\sqb{\mathbf{x}},\sqb{\mathbf{p}}}} \cdot
  	\tfrac{\sqb{s}}{\sqb{s}^2 + \tilde{\epsilon}} \cdot \sqb{\dot{V}^{\dbr{\mathrm{B}}}} }
  	\enspace.
  \end{equation}
  Similarly, interval-based generalizations can be defined for all before-mentioned second-order formulations. For a detailed discussion of restrictions in the case that accounts for integrator extensions in the definition of the sliding surface $s=0$, the reader is referred to~\cite{Rauh_Senkel_springer_2015b}.


  In previous work such as \cite{RAUH20176283,7739793,Rauh_Rauh:2015_TIE,Rauh_Senkel_springer_2015b}, where this approach was applied successfully to the control of the thermal behavior of high-temperature solid oxide fuel cells as well as for the control of inverted pendulum systems, the {\sc{C++}} toolbox {\sc{C-XSC}}~\cite{Rauh_XSC_URL} was used. Required derivatives, necessary to transform general nonlinear state equations into the system representation assumed in this section, can easily be obtained with the help of algorithmic differentiation{, cf.~\cite{Rauh_Senkel_springer_2015}}. The template-based library \textsc{FADBAD++} has shown its efficiency for this purpose, because it cannot only be applied to expressions involving classical point-valued data types but also interval variables. 
  
  It has to be pointed out that the actual implementation of the control laws~(\ref{Rauh_eq:u_interval1})--(\ref{Rauh_eq:u_interval1b}) is performed in such a way that a point-valued control signal $v$ is chosen from the computed intervals (\ref{Rauh_eq:u_interval1})--(\ref{Rauh_eq:u_interval1b}) so that it guarantees asymptotic stability regardless of the sign of ${b}({\mathbf{x}},{\mathbf{p}})$. According to~\cite{Rauh_Senkel_springer_2015b}, this is done by testing the negative definiteness of the Lyapunov function candidate for the infima and suprema  $\ul{v}:=\mathrm{inf}\{\sqb{v}\}$ and $\ol{v}:=\mathrm{sup}\{\sqb{v}\}$. To account for roundoff and representation errors, these values are inflated by a small constant $\epsilon>0$ to obtain the final set of candidates 
  $
  \mathcal{V}:= 
  \left\{
  \ul{v}-\epsilon,
  \ul{v}+\epsilon,
  \ol{v}-\epsilon,
  \ol{v}+\epsilon
  \right\}
  $, 
  from which the signal with minimum absolute value is chosen that ensures $\dot{V} < 0$ (or more generally $\dot{V}^{\dbr{j,\mathrm{A}}}<0$, $\dot{V}^{\dbr{j,\mathrm{B}}}<0$, $j \in \{\mathrm{I},\mathrm{II}\}$) despite the considered interval uncertainty {as a criterion for stability of the uncertain dynamic system model}.

\section{Guaranteed Model Predictive Control}\label{sec:mpc}
\label{sec2}

\subsection{{Goal of the Control Synthesis}}\label{sec:pendulum}
 {Predictive control techniques~\cite{camacho2013model}} are powerful approaches for controlling uncertain systems, where the corresponding dynamics are formulated {in terms of} initial value problems for finite-dimensional sets of nonlinear ordinary differential equations. Besides stabilizing the dynamics towards a desired reference trajectory, it is possible to formulate the involved optimization problem in such a way that constraints for the admissible state trajectories and system inputs are handled. Classical approaches for nonlinear model predictive control commonly do not account for interval bounds in initial states and system parameters. 
 
\begin{figure}[htp]
	\begin{center}
		\includegraphics[width=\linewidth]{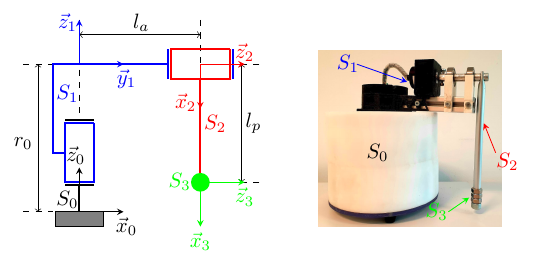}
		\caption{{Application scenario: Swing-up of an inverted pendulum according to~\cite{Fnadi2023}.}}
		\label{NMPC1} 
	\end{center}
\end{figure}

 {The main drawback of neglecting this information about uncertainty is the lack of a robustness measure of the resulting control. This neglection may lead to scenarios where the determined control signal does only stabilize the system model for parts of the possible parameter ranges, while the resulting dynamics are unstable for others. This is especially critical for open-loop unstable system models such as the inverted pendulum system in Figure~\ref{NMPC1}, for which a model predictive controller was designed in~\cite{Fnadi2023}. This controller simultaneously aims at a swing-up of the pendulum arm $S_2$ by rotation of the arm $S_1$ and holding $S_2$ in its unstable upright equilibrium by means of pure inertia coupling.}
 
 \subsection{{Formulation of a Set-Valued Initial Value Problem for Model Predictive Control}}
 {The issue of a missing robustness quantification of classical model predictive controllers} is resolved in this section, by presenting a control procedure for interval-valued initial value problems in the form~\cite{Fnadi2023}
 \begin{eqnarray}\label{ODE}
 	\left\{\begin{array}{l}
 		{\dot{\mathbf{x}}(t)=\mathbf{f}(t, \mathbf{x}(t),\mathbf{u}(t), \mathbf{p})} \\
 		\mathbf{x}_0 \in\left[\mathbf{{x}}_{0}\right] \subseteq {\mathbb{IR}}^{n}
 		\\
 		{\mathbf{u}(t) \in\left[\mathbf{{u}}\right] \subseteq {\mathbb{IR}^m}}
 		\\
 		\mathbf{p} \in\left[\mathbf{{p}}\right] \subseteq {\mathbb{IR}^p},
 	\end{array}\right.
 \end{eqnarray}
 where the state vector is denoted by $\mathbf{x}(t)$, the vector of dynamic parameters by $\mathbf{p}$, and the control vector by~$\mathbf{u}$. The sets $\left[\mathbf{{x}}_{0}\right]=\begin{bmatrix}
 	\left[{{x}}_{10}\right] & \ldots &\left[{{x}}_{n0}\right]
 \end{bmatrix}^T$, $\left[\mathbf{{u}}_{}\right]=\begin{bmatrix}
 	\left[{{u}}_{1}\right] & \ldots & \left[{{u}}_{m}\right]
 \end{bmatrix}^T$, and $\left[\mathbf{{p}}_{}\right]=
 \begin{bmatrix}
 	\left[{{p}}_{1}\right] & \ldots & \left[{{p}}_\mathrm{p}\right]
 \end{bmatrix}^T$, expressed as interval boxes, are respectively the initial condition of the state vector, the interval-bounded input {(assumed without loss of generality to have constant bounds over the time horizon of interest)}, and the set of feasible dynamic parameters. We assume that this interval-based initial value problem has a unique solution {$\mathbf{x}(t,\mathbf{x}_0,\mathbf{u},\mathbf{p})$} at $t>0$. {A sufficient condition for this is that $\mathbf{f}: \mathbb{R} \times \mathbb{R}^{n}\times \mathbb{R}^m\times \mathbb{R}^p \rightarrow \mathbb{R}^{n}$ is continuous in $t$ and \mbox{Lipschitz} in $\mathbf{x}(t)$.} For the use of the proposed solution algorithm, we further require that $\mathbf{f}$ is sufficiently smooth, i.e., of class $C^k$. For an approach applicable to linear time-invariant and linear parameter-varying systems, which is similar to the one described subsequently, the reader is referred to \cite{DOSREISDESOUZA2022109951,Reis,9496205}.

\subsection{Interval-Based Nonlinear Model Predictive Control}

The interval-based nonlinear model predictive control approach presented in \cite{Fnadi2023,fnadi2021experimental} is a generalization of the optimization approach in \cite{erice:2007} and is based on the computation of interval bounds for the control sequence over a receding horizon which takes into account bounded uncertainties in the parameters of the dynamic system model and in the measured data. 

The control intervals are calculated such that the {state trajectories converge (monotonically)} to the set-point interval {over time} (i.e., $\textbf{x}(t_j)\to[\textbf{x}_\mathrm{r}]$, as $j \to \infty$, and all the state and input constraints are satisfied (i.e., $\textbf{x}(t_j)\in[\textbf{x}(t_j)]$ and $\textbf{u}(t_j)\in[\textbf{u}(t_j)], ~ ~ \forall j$). {Here, the set-point interval either represents a bounded domain around a desired operating point (in the pendulum example, the upright position of the arm $S_2$) or a bounded tube around a desired time-dependent trajectory.}

{The interval algorithm for determining the model predictive control strategy} encompasses two stages~\cite{fnadi2021experimental}, namely,
\begin{itemize}
	\item  \textbf{Filtering and branching:} This first step provides a sequence of guaranteed input interval boxes at each time-step
	$t_k$ over the prediction horizon $N_\mathrm{p}$, denoted as $[\mathbf{U}(t_k)] = [\mathbf{{u}}(t_{k})] \times [\mathbf{{u}}{(t_{k+1})}] \times \ldots \times [\mathbf{{u}}(t_{k+N_\mathrm{p} -1})]$. Branching and filtering procedures allow the computation of safe input intervals\footnote{{For the implementation of this branching algorithm we assume piecewise constant input intervals. Therefore, the more general notation $[\mathbf{{u}}(t_{k})]$ is subsequently replaced by $[\mathbf{{u}}_k]$ to indicate control signals only changing their value at the points in time $t_k$.}} along the receding time horizon. {In this context, safety means that the state constraints that are specified in the form of admissible intervals in each time step} (i.e., $\forall j, \left[\mathbf{{x}}(t_{j})\right] \subseteq [\mathbf{x}_{\min}(t_j), \mathbf{x}_{\max}(t_j)]$ {are guaranteed to be obeyed despite the parameter uncertainty}, where $\mathbf{x}_{\min}(t_j)$ and $\mathbf{x}_{\max}(t_j)$ are the bounds for the admissible domain for each state variable). {In addition, the desired} convergence to the reference interval (i.e., $[\mathbf{x}_j] \rightarrow [\mathbf{x}_{\mathrm{r}}]$) {needs to be ensured}. 
	
	\item  \textbf{Interval optimization:}  Since safe inputs are computed over a finite time horizon, the optimization algorithm is launched to compute the optimal inputs $[\mathbf{U}(t_k)]^\star$ by minimizing as much as possible a newly formulated interval objective function to reduce both the norm of the input intervals {(i.e., to reduce the control effort)} and the error between the predicted {and} the reference {set-point (or, resp., reference trajectory)}. 	
\end{itemize}

\subsection{Formulation and Minimization of Interval Cost Functions}
\label{Cost}

{To determine} feasible intervals of input values that satisfy the constraints on the state trajectories and system inputs, an optimization procedure is required. {It aims at finding} the optimal control box and --- on this basis --- a point-valued input $\mathbf{u}(t_k)$ that can be applied to the actuator of the considered system \cite{Fnadi2023,fnadi2021experimental}. The continuous cost function to be minimized can be expressed over the prediction horizon $T_\mathrm{p}=T_\mathrm{c}\times N_\mathrm{p}$, with the control update step size $T_\mathrm{c}$, as 
\begin{equation}\label{of}
	J(\mathbf{x}(t), \mathbf{{u}}(t))=\int_{t}^{t+T_\mathrm{p}} F(\mathbf{x}(\tau), \mathbf{u}(\tau)) \mathrm{d} \tau \enspace .
\end{equation}
Here, $F$ is commonly chosen as a quadratic function that {depends on} the norm of the inputs and the error between the predicted {states (resp, outputs)} $\mathbf{x}(t)$ and the reference $\mathbf{x}_\mathrm{r}$ according to
\begin{equation}
	{F(\mathbf{x}(t), \mathbf{u}(\tau))=\left(\mathbf{x}(t)-\mathbf{x}_\mathrm{r}\right)^{T} \mathbf{Q}\left(\mathbf{x}(t)-\mathbf{x}_\mathrm{r}\right)+\mathbf{u}(t)^{T} \mathbf{R} {\mathbf{u}}(t) \enspace ,}
\end{equation}
where $\mathbf{Q}$ and $\mathbf{R}$ are both positive (semi-)definite weighting matrices. 

Under the assumption of piecewise constant control inputs, the continuous objective function (\ref{of}) takes the form
\begin{small}
	\begin{equation}\label{op2}
		\begin{split}
			J(\mathbf{x}(t), {\mathbf{u}}(t)) &= \int_{t}^{t+T_\mathrm{p}}\left(\left(\mathbf{x}(\tau)-\mathbf{x}_\mathrm{r}\right)^{T} \mathbf{Q}\left(\mathbf{x}(\tau)-\mathbf{x}_\mathrm{r}\right)+\mathbf{u}(\tau)^{T} \mathbf{R} {\mathbf{u}}(\tau)\right) \mathrm{d} \tau\\
			&=   \sum_{k=1}^{N_\mathrm{p}} \int_{t+(k-1) T_\mathrm{c}}^{t+k T_\mathrm{c}}\left(\left(\mathbf{x}(\tau)-\mathbf{x}_\mathrm{r}\right)^{T} \mathbf{Q}\left(\mathbf{x}(\tau)-\mathbf{x}_\mathrm{r}\right)\right) \mathrm{d} \tau+ \sum_{k=1}^{N_\mathrm{p}} \int_{t+(k-1) T_\mathrm{c}}^{t+k T_\mathrm{c}} \mathbf{u}(\tau)^{T} \mathbf{R} \mathbf{u}(\tau) \mathrm{d} \tau \enspace .
		\end{split}
		\raisetag{19mm}
\end{equation}\end{small}

As shown in \cite{Fnadi2023,fnadi2021experimental}, the value of cost function (\ref{op2}) can be enclosed with the help of techniques from interval analysis. This is achieved by applying a validated integration method for the initial value problem under consideration. This method provides a list of tight enclosures $[{\mathbf{x}_0}],[{\mathbf{x}_1}],\ldots,[\mathbf{x}_{K}]$ for the solution of the model (\ref{ODE}) {on each interval between two subsequent time steps}. These enclosures are obtained starting from the initial conditions $[\mathbf{{x}}_0]$ with bounded, piecewise constant control sequences $[\mathbf{{u}}_j]$ {so that the obtained boxes satisfy the condition} $\mathbf{x}(t)\in [\mathbf{{x}}_t]$ {for all} $t\in [t+(k-1)T_\mathrm{c}\; ; \; t+kT_\mathrm{c}]$. By using these {piecewise defined} interval enclosures, bounds for the cost function are obtained according to
\begin{equation} \label{opp}
	\begin{aligned}
		J(\mathbf{x}_t, \mathbf{u}) &\in 
		T_\mathrm{c}  \sum_{j=k}^{k+N_\mathrm{p}-1} \left(\left([\mathbf{{x}}_{[j:j+1]}]-\mathbf{x}_\mathrm{r}\right)^{T} \mathbf{Q}\left([\mathbf{{x}}_{[j:j+1]}]-\mathbf{x}_\mathrm{r}\right) ~+~ [\mathbf{u}_{j}]^{T} \mathbf{R} [\mathbf{u}_{j}] \right)
		\\
		&\leq  \mathrm{sup}\left\{  T_\mathrm{c}  \sum_{j=k}^{k+N_\mathrm{p}-1} \left(\left([\mathbf{{x}}_{[j:j+1]}]-\mathbf{x}_\mathrm{r}\right)^{T} \mathbf{Q}\left([\mathbf{{x}}_{[j:j+1]}]-\mathbf{x}_\mathrm{r}\right) ~+~ [\mathbf{u}_{j}]^{T} \mathbf{R} [\mathbf{u}_{j}] \right)\right\} \enspace,
	\end{aligned}
\end{equation}
{where $[\mathbf{{x}}_{[j:j+1]}]$ is a guaranteed enclosure of all reachable states over the temporal slice $t \in [t_j\; ; \; t_j + T_\mathrm{c}]$.}

\begin{figure}[htp]
	\begin{center}
		\includegraphics[width=\linewidth]{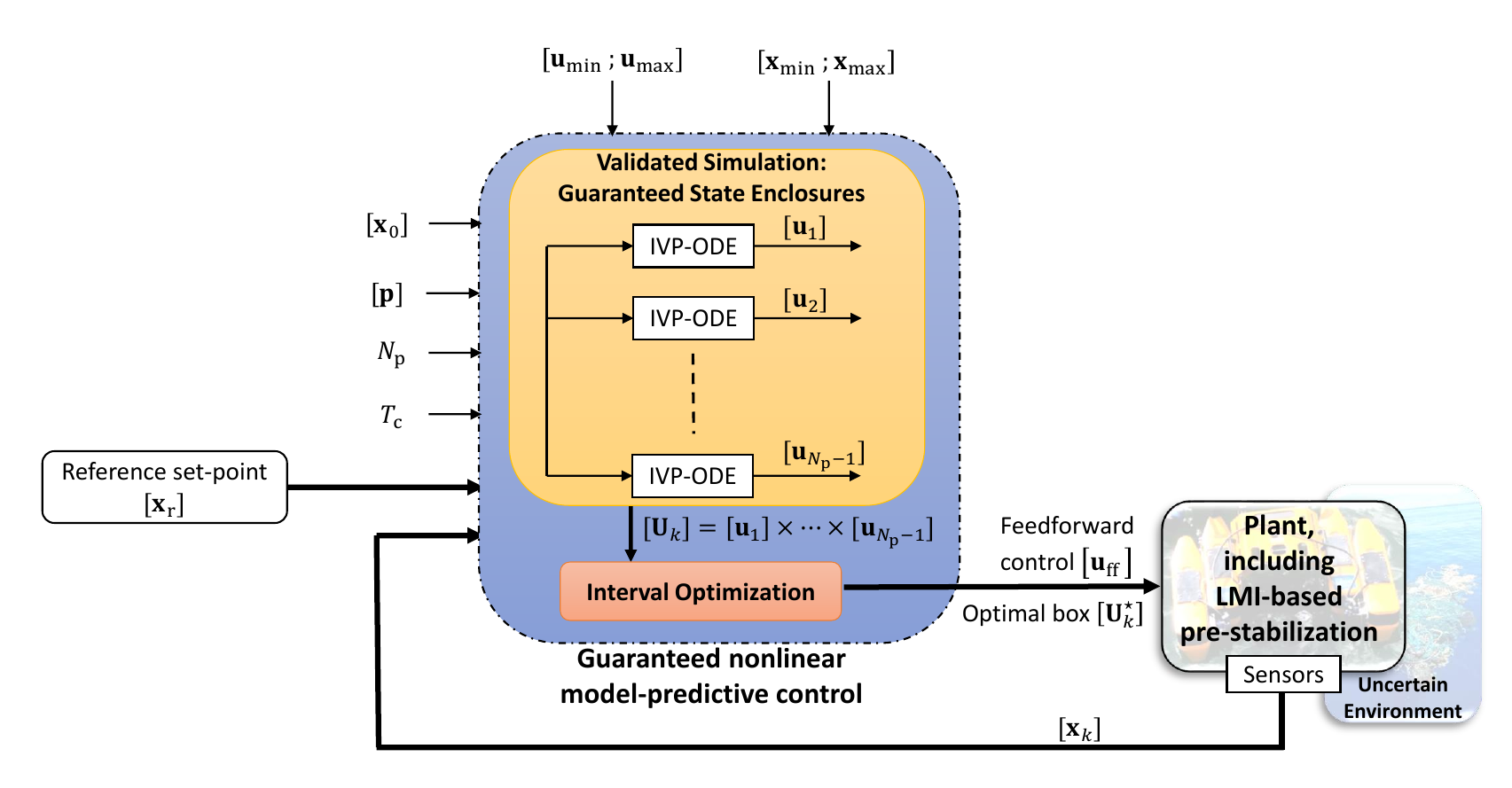}
		\caption{Overall structure of the interval-based nonlinear model predictive control approach, extended by an underlying pre-stabilization of the plant dynamics.}
		\label{NMPC} 
	\end{center}
\end{figure}

Figure~\ref{NMPC} gives an overview of {a possible} extension of the interval-based nonlinear model predictive control procedure. {This extension is represented by a robust pre-stabilization} of open-loop unstable plants with the help of linear matrix inequality (LMI) techniques. In such cases, the interval-based optimization procedure does not directly provide the overall control sequence but rather a feedforward control action $[\mathbf{u}_\mathrm{ff}]$ that allows for optimizing the tracking behavior within the operating domain in which the underlying feedback controller possesses provable asymptotic stability properties~\cite{Fnadi2023}. {For an application of this control procedure to the swing-up control of a rotary inverse pendulum sketched in Sec.~\ref{sec:pendulum}, the reader is referred to~\cite{Fnadi2023}.}

\section{{Interval Observer-Based Identification of State-Dependent Static Subsystem Models}} \label{sec:IntervalObserverLR}

\subsection{{Goal of the Estimator Synthesis}}
{Set-based estimation strategies can be used beyond the case of reconstructing internal system states in a model-based manner by means of a knowledge about control variables and measured outputs. If combined efficiently with contractor-based procedures, it is also possible to exploit the estimates provided by an interval observer to determine bounds for nonlinear static subsystem models that are embedded in the overall system representation. An example for this is the reconstruction of the nonlinear open-circuit voltage characteristic of a Lithium-ion battery on the basis of an electric equivalent circuit model, knowledge about the terminal current as an actuation signal, and the measurement of the terminal voltage with bounded uncertainty \cite{Lahme2023,Lahme_mmar_2023}.}



{For this estimation task, consider the}
equivalent circuit model of the battery as described in \cite{Erdinc.2009}. {It is represented by the quasi-linear state equations}
%
\begin{align}\label{eq:StateEqLR}
	\dot{\mathbf{x}}(t) &= {\mathbf{A}}\rb{\sigma(t)}\cdot {\mathbf{x}}(t) + {\mathbf{b}}\rb{\sigma(t)} \cdot u(t)\\
	&=\begin{bmatrix}
		0  &   0                  & 0\\[1mm]
		0  &  \frac{-1}{{C}_\mathrm{TS}\rb{\sigma(t)}\cdot{R}_\mathrm{TS}\rb{\sigma(t)}} & 0\\[1mm]
		0  &   0                  & \frac{-1}{{C}_\mathrm{TL}\rb{\sigma(t)}\cdot{R}_\mathrm{TL}\rb{\sigma(t)}}\\[1mm]
	\end{bmatrix}
	\cdot
	{\textbf{x}}(t)
	+
	\begin{bmatrix}
		\frac{-1}{C_\mathrm{Bat}} \\[1mm] \frac{1}{C_\mathrm{TS}\rb{\sigma(t)}} \\[1mm]  \frac{1}{C_\mathrm{TL}\rb{\sigma(t)}}
	\end{bmatrix}
	\cdot
	u(t)
	\nonumber  
\end{align}
with the state vector 
\begin{equation}\label{eq:state_vec_batt}
	{\mathbf{x}}(t) = 
	\begin{bmatrix}
		\sigma(t) & {v}_\mathrm{TS}(t) & {v}_\mathrm{TL}(t)
	\end{bmatrix}^T
	\enspace.
\end{equation}
These state variables denote the state of charge $\sigma$ of the battery as well as two voltages ${v}_\mathrm{TS}$ and ${v}_\mathrm{TL}$ across resistor--capacitor sub-networks included in the equivalent circuit to represent delay phenomena between {step}-like changes of the terminal current and measurable variations of the terminal voltage. According to \cite{Lahme2023}, interval enclosures of the state variables are estimated after reformulation of both the state equations~(\ref{eq:StateEqLR})
and the output equation
\begin{align}			
	\tilde{v}_\mathrm{OC}(\sigma(t))&={\mathbf{C}}\rb{\sigma(t)}\cdot {\mathbf{x}}(t)\\
	& =\eta_\mathrm{OC}\rb{\sigma(t)} \cdot \sigma(t)
	= {v}_\mathrm{OC}(\sigma(t)) - v_0 - v_2
	\\
	&= \rb{v_0\cdot\frac{ e^{v_1\cdot \sigma(t) } -1 }{\sigma(t)} + v_3 
		+ v_4\cdot \sigma(t) + v_5\cdot {\rb{\sigma(t)}^2}} \cdot \sigma(t) 
	\label{eq:VOCQuasiLin}
\end{align}
in a quasi-linear form in which state of charge dependent matrices and vectors ${\mathbf{A}}\rb{\sigma(t)}$, ${\mathbf{b}}\rb{\sigma(t)}$, and ${\mathbf{C}}\rb{\sigma(t)}$, respectively, are separated from the state vector ${\mathbf{x}}(t)$ and the input variable $u(t)$, which is represented by the terminal current $i_\mathrm{T}$. {Nonlinear approximations for the state of charge dependent variations of the resistances and capacitances 
${R}_\mathrm{TS}\rb{\sigma(t)}$, ${R}_\mathrm{TL}\rb{\sigma(t)}$, ${C}_\mathrm{TS}\rb{\sigma(t)}$, and ${C}_\mathrm{TL}\rb{\sigma(t)}$ for processes with short and long time constants (indices $\mathrm{TS}$ and $\mathrm{TL}$, respectively) and nominal values for the coefficients $v_0,\ldots,v_5$ of the open-circuit voltage ${v}_\mathrm{OC}(\sigma(t))$ are identified beforehand by means of experimental data according to \cite{reuter2016}. A conservative set-based enclosure of the resulting matrices then forms the basis for the subsequent observer interval observer design.}

\subsection{{Interval Observer Synthesis}}

For the {design of an interval observer}, cooperativity of the system model (as discussed in the introduction of this article) as well as stability of the estimation error dynamics {need to be} ensured. {This can be achieved} by choosing the observer gain matrix $\mathbf{H}$ according to \cite{Hildebrandt.2020}, in the form
\begin{equation}\label{eq:gain_H}
	{\mathbf{H}}
	=
	\left[
	\begin{array}{ccc}
		h_1 & 0 & 0 \\		
	\end{array}
	\right]^T
\end{equation}
with the constant value $h_1 > 0$.

In such a way, interval bounds $\hat{\mathbf{x}} \in \sqb{\hat{\mathbf{x}}} := \intv{\hat{\mathbf{\ul{x}}}}{\hat{\mathbf{\ol{x}}}}$ can be determined for the true, non-measurable state vector {${\mathbf{x}}(t)$ defined in~(\ref{eq:state_vec_batt})} 
by {means of} lower and upper bounding systems \cite{Hildebrandt.2020,raissi_at}, following Müller's theorem \cite{mueller:1927}. {These bounding systems are given by}
\begin{equation}\label{eq:IntervalObserverLR}
	\ul{{\mathbf{A}}_\mathrm{O}  \hat{\mathbf{{x}}}} + \ul{\mathbf{b} \mathbf{u}} + \mathbf{H} \ul{\mathbf{y}}_\mathrm{m}
	\le
	\dot{\hat{\mathbf{x}}}
	\le 
	\ol{{\mathbf{A}}_\mathrm{O} \hat{\mathbf{{x}}}} + \ol{\mathbf{b} \mathbf{u}} + \mathbf{H} \ol{\mathbf{y}}_\mathrm{m} 
	\enspace ,~
	\hat{\mathbf{x}} \in \sqb{\hat{\mathbf{x}}}\enspace ,
\end{equation}
where
\begin{equation}
	{\mathbf{A}}_\mathrm{O}
	=
	{\mathbf{A}} - \mathbf{H} \mathbf{C}
	\enspace
	\in
	\enspace
	\intv{\ul{\mathbf{A}}_\mathrm{O}}{\ol{\mathbf{A}}_\mathrm{O}}
\end{equation}
is a Metzler matrix \cite{Gennat.2008,Kersten.2019,Kaczorek} with non-negative off-diagonal elements
and 
\begin{equation}\label{eq:MeasurementIntervalLR}
	\sqb{{\mathbf{y}}_\mathrm{m}} := 
	\intv{\ul{\mathbf{y}}_\mathrm{m} }{\ol{\mathbf{y}}_\mathrm{m} }
	= {\mathbf{y}}_\mathrm{m} + \intv{-\Delta{\mathbf{y}}_\mathrm{m}}{\Delta{\mathbf{y}}_\mathrm{m}}
\end{equation}
is the measured system output with bounded uncertainty. {To remove the structural limitation for the gain matrix~(\ref{eq:gain_H}) in which two of the elements need to be strictly equal to zero, an extended as well as cascaded observer design was recently published in~\cite{Lahme_ECC_2024}. This extension provides additional tuning parameters for the observer design so that the estimation uncertainty is typically reduced by a reduction of the sensitivity of the state bounds against model inaccuracies and measurement errors.}

{The outcome of the observer-based state reconstruction is used in the following subsection for} an interval-based set intersection approach. {This procedure allows for identifying} the characteristic of the open circuit voltage ${v}_\mathrm{OC}(\sigma(t))$ {which was structurally unknown to the interval observer and only considered by an outer interval enclosure for the matrix ${\mathbf{C}}\rb{\sigma(t)}$.}

\subsection{{Implementation for the Estimation of the Open-Circuit Voltage Characteristic of a Lithium-Ion Battery}}

\begin{figure}[t]
	\begin{center}
		\includegraphics[width=\linewidth]{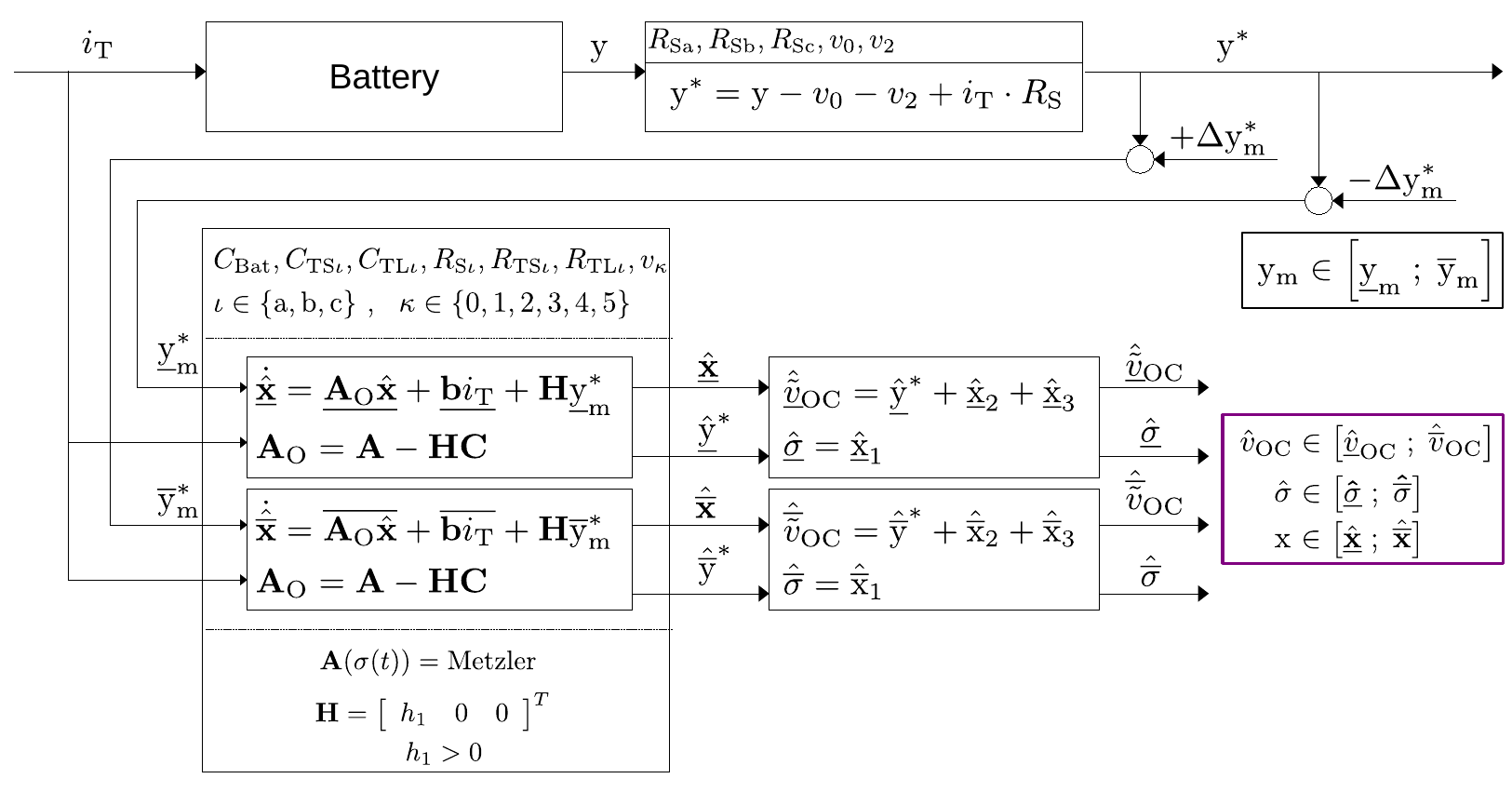}
	\end{center}
	\caption{Estimation of the open-circuit voltage and the state of charge.\label{fig:EstimationSchemeLR}}		
\end{figure}

{According to the equivalent circuit representation of the dynamics of a Lithium-ion battery, an interval-based  estimate can be determined for the open-circuit voltage according to}
\begin{equation}\label{eq:out_batt}
	\tilde{v}_\mathrm{OC}(t) = \mathrm{y}^*_\mathrm{m}(t)  + \hat{v}_\mathrm{TS}(t) + \hat{v}_\mathrm{TL}(t)
	\enspace ,
\end{equation}	
with $\hat{v}_\mathrm{TS}(t) \in \sqb{\hat{v}_\mathrm{TS}(t)}$ and $\hat{v}_\mathrm{TL}(t) \in \sqb{\hat{v}_\mathrm{TL}(t)}$. {Together with the interval enclosure $\sqb{\hat{\sigma}}$ for the state of charge, an} interval box $\sqb{\mathbf{\Gamma}} = \sqb{\sigma(t)} \times \sqb{\tilde{v}_\mathrm{OC}(t)}$, $\sigma(t) \in \intv{\ul{\sigma}(t)}{\ol{\sigma}(t)}$ and $\tilde{v}_\mathrm{OC}(t) \in \intv{\ul{\tilde{v}}_\mathrm{OC}(t)}{\ol{\tilde{v}}_\mathrm{OC}(t)}$, {can be constructed}. {All required computational steps are summarized in the block diagram depicted in Fig.~\ref{fig:EstimationSchemeLR}.}

\begin{figure}[htp]
	\begin{center}
			\includegraphics[width=\linewidth]{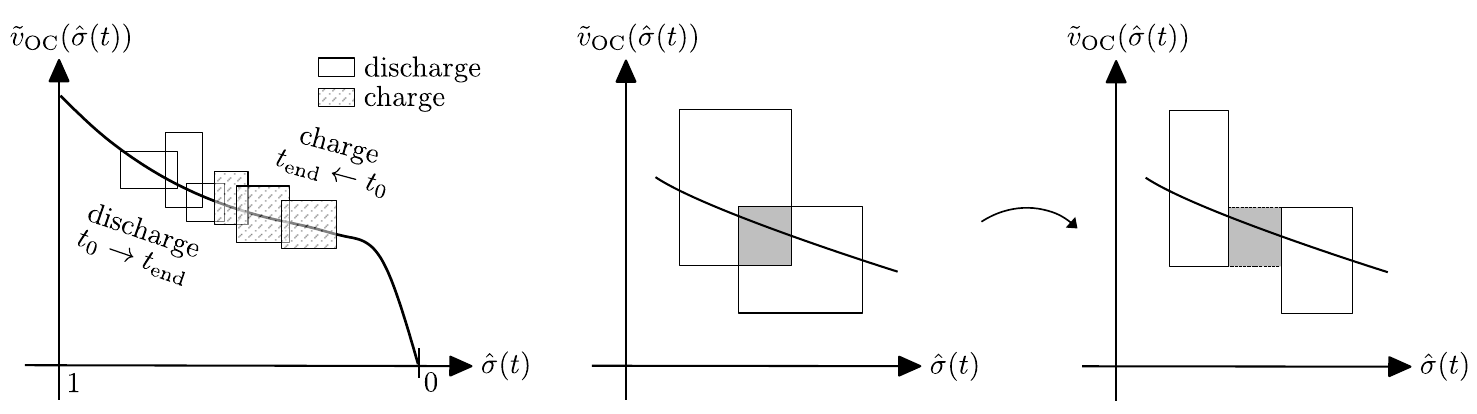}
		\caption{Identification of nonlinear dependencies using interval methods.}
		\label{fig:IdentificationLR}
	\end{center}
\end{figure}

Successive intersections of these interval boxes {$\sqb{\mathbf{\Gamma}}$} --- at different points in time --- are utilized to improve the approximation of the true ${v}_\mathrm{OC}(\sigma(t))$ characteristic according to Figure~\ref{fig:IdentificationLR}.

For strategies, allowing for a reduction of the computational complexity due to a large number of interval boxes to be intersected, the interval merging routine according to \cite{Irina} is employed. Moreover, refinements of the estimated open-circuit voltage characteristic are possible by implementing a forward--backward contractor, as discussed from a methodological point of view in \cite{AppliedIntervalAnalysis}. These refinements have been demonstrated recently in \cite{Lahme_mmar_2023}.

 \section{Combination of Set-Based and Stochastic Uncertainty Representations: An Application to Iterative-Learning Observer Design}\label{sec:ellipse}
 \subsection{{Goal of the Estimator Synthesis}}
 {Many practical applications are characterized by the fact that uncertainties and disturbances are neither characterized purely by set-based nor by stochastic representations. Then, it is desired to have a unified  representation of uncertainty that can be employed to implement state estimators. In addition, many real-life applications are also subject to nonlinearities, making an exact solution of the estimation task impossible. Widely used approximation techniques such as the Extended or Unscented Kalman Filter \cite{Julier:2000,Stengel:2001,Rauh_Anderson} have the disadvantage that they cannot take into consideration the worst-case influence of linearization errors of both nonlinear state equations and measurement models.}
 
 Therefore, this section introduces a combination of stochastic and set-based uncertainty representations for the estimator design described in Sec.~\ref{sec:problem}. This approach allows, on the one hand, for a rigorous quantification of predefined confidence levels in stochastic state estimation procedures. On the other hand, it allows for handling nonlinearities in such a way that the previously mentioned tolerance bounds are definitely not determined in an overly optimistic manner \cite{Rauh_ECC_2022,9841329,math10163011}. Inspired by normally distributed uncertainty models, the suggested approach is based on the use of an ellipsoidal calculus.
 
 Practical application scenarios are, for example, the state estimation of marine vessels on the basis of GPS position measurements~\cite{math10163011}, state estimation for battery systems~\cite{Rauh_Jaulin}, or state estimators that iteratively enhance the estimation accuracy for system models that are characterized by periodically recurring trajectories~\cite{Rauh_ECC_2022}. An example for the latter is given in~\cite{9841329} which demonstrates the successful application of the observer for the state and disturbance estimation of a Lithium-ion battery with periodically recurring input current profiles.
 
 \subsection{{Preliminaries for Ellipsoidal State Enclosures}}
{In the following, we focus on an estimator synthesis in a mixed stochastic and set-based uncertainty representation} for the case of periodically recurring trajectories and disturbance profiles (e.g., due to errors caused by a deterministic model mismatch due to purposefully performed model simplifications). For that purpose, assume the discrete time system model
  \begin{equation}\label{eq:sys_QL}
  	\mathbf{z}_{k+1} = \boldsymbol{\Phi}\rb{\mathbf{z}_k,\mathbf{p}}\cdot \mathbf{z}_k
  \end{equation}
 with the bounded parameter vector $\mathbf{p} \in \intv{\ul{\mathbf{p}}}{\ol{\mathbf{p}}}$. As discussed in the following, the vector $\mathbf{z}_k \in \mathbb{R}^{\tilde{n}}$ consists of both system states and noise variables. Furthermore, assume that the ellipsoidal domain
  \begin{equation}\label{eq:ellips}
  		{\mathcal{E}}_k\rb{\boldsymbol{\mu}_k, \boldsymbol{\Gamma}_k^\prime, r}\\
  		:=\left\{\mathbf{z}_k\in \mathbb{R}^n \bigm\vert
  		\rb{\mathbf{z}_k-\boldsymbol{\mu}_k}^T {\boldsymbol{\Gamma}_k^\prime}^{-T} {\boldsymbol{\Gamma}_k^\prime}^{-1} \rb{\mathbf{z}_k-\boldsymbol{\mu}_k} \le r^2
  		\right\}
  \end{equation}
  specifies the confidence bound of a given percentage if the vector $\mathbf{z}_k$ is normally distributed with the
  {ellipsoid midpoint as the expected value \mbox{$\boldsymbol{\mu}_k \in \mathbb{R}^{\tilde{n}}$} and the} positive definite covariance matrix $\mathbf{Q}_k^\prime = \boldsymbol{\Gamma}_k^\prime {\boldsymbol{\Gamma}_k^{\prime}}^T \succ 0$. {In~(\ref{eq:ellips})}, the parameter $r$ describes a magnification factor according to~\cite{Wang:2015} associated with {the before-mentioned} confidence level.

 \subsection{Ellipsoidal Calculus for the Guaranteed Preduction of Covariance Ellipsoids} \label{sec:predict_ellipsoid}
  For the implementation of an ellipsoidal covariance prediction step (based on the references~\cite{a14030088,Rauh_Jaulin_ICCAD,Rauh_Jaulin}), we use the definitions
  \begin{equation}
  	\boldsymbol{\Gamma}_k := r\cdot \boldsymbol{\Gamma}_k^\prime
  	\quad \text{and} \quad
  	\mathbf{Q}_k := r^2 \cdot \mathbf{Q}_k^\prime
  \end{equation}
  {transferring the confidence ellipsoid~(\ref{eq:ellips}) into the normalized standard form
  \begin{equation}
  	{\mathcal{E}}_k^{\boldsymbol{\Gamma}}\rb{\boldsymbol{\mu}_k, \boldsymbol{\Gamma}_k}\\
  	:=\left\{\mathbf{z}_k\in \mathbb{R}^n \bigm\vert
  	\rb{\mathbf{z}_k-\boldsymbol{\mu}_k}^T \boldsymbol{\Gamma}_k^{-T} \boldsymbol{\Gamma}_k^{-1} \rb{\mathbf{z}_k-\boldsymbol{\mu}_k} \le 1
  	\right\}
  	\enspace .
  \end{equation}
  }
   
  For the compactness of notation, reformulate the system model~(\ref{eq:sys_QL}) into the form
  \begin{equation}\label{eq:split}
  	\mathbf{z}_{k+1} = 
  	\boldsymbol{\Phi}\rb{\mathbf{z}_{k},\mathbf{p}}\cdot \check{\mathbf{z}}_k + 
  	\tilde{\boldsymbol{\Phi}} \cdot \boldsymbol{\mu}_k + \rb{\boldsymbol{\Phi}\rb{\mathbf{z}_{k},\mathbf{p}} - \tilde{\boldsymbol{\Phi}}} \cdot \boldsymbol{\mu}_k 
  \end{equation}
  with $\mathbf{z}_{k} = \check{\mathbf{z}}_k + \boldsymbol{\mu}_k$, where
  \begin{equation}
  	\mathbf{z}_{k} \in 
  	{\mathcal{E}}_k = {\mathcal{E}}_k\rb{\boldsymbol{\mu}_k, \boldsymbol{\Gamma}_k^\prime ,r}\enspace .\label{eq:thick_ell_complete}
  \end{equation}
  Here, ${\mathcal{E}}_k$ denotes the uncertainty on the non-origin centered states $\mathbf{z}_k$,
  \begin{equation}	
  	\check{\mathbf{z}}_k \in \check{{\mathcal{E}}}_k =\check{{\mathcal{E}}}_k\rb{\mathbf{0}, \boldsymbol{\Gamma}_k^\prime ,r}\label{eq:thick_ell_center0}
  \end{equation}
  the uncertainty on $\check{\mathbf{z}}_k$ after shifting the ellipsoid to the origin, and
  \begin{equation}
  	\tilde{\boldsymbol{\Phi}} = \boldsymbol{\Phi}\rb{\boldsymbol{\mu}_k,\mathrm{mid}\rb{\sqb{\mathbf{p}}}}\label{eq:A_tilde}
  \end{equation}
  is the midpoint approximation of the quasi-linear system matrix with
  \begin{equation}
  	\mathrm{mid}\rb{\sqb{\mathbf{p}}} = \frac{1}{2}\cdot \rb{\ul{\mathbf{p}}+\ol{\mathbf{p}}}\label{eq:par_box}
  	\enspace .
  \end{equation}
  
  Let also $\square{\mathcal{E}_k}$ denote an axis-aligned enclosure of {the ellipsoid} ${\mathcal{E}}_k$ {defined in~(\ref{eq:ellips})} in the form of an $\tilde{n}$-dimensional interval box. An alternative to the definition~(\ref{eq:A_tilde}) is given by
  \begin{equation}\label{eq:A_tilde1}
  	\tilde{\boldsymbol{\Phi}} = \mathrm{mid}\rb{\boldsymbol{\Phi}\rb{\square{\mathcal{E}_k},\sqb{\mathbf{p}}}}
  	\enspace ,
  \end{equation}
  which is preferable if (\ref{eq:A_tilde}) and~(\ref{eq:A_tilde1}) strongly differ from each other in the case of large uncertainty.
  
  Then, a confidence bound of the predicted states $\mathbf{z}_{k+1}$ {corresponding to a given confidence level with the associated} magnification $r$ {leads to} the ellipsoid ${\mathcal{E}}_{k+1}\rb{\boldsymbol{\mu}_{k+1}, \boldsymbol{\Gamma}_{k+1}^\prime,r}$ with the covariance $\mathbf{Q}_{k+1}^\prime= \boldsymbol{\Gamma}_{k+1}^\prime \rb{{\boldsymbol{\Gamma}_{k+1}^{\prime}}}^{T}$ for which the parameters are computed in the following steps. For further details of the following algorithm, Consisting of the prediction steps \textbf{P1}--\textbf{P3}, the reader is referred to the reference~\cite{Rauh_ECC_2022}.

  \begin{description}
  	\item[\textbf{P1}] Apply
  	\begin{equation}\label{eq:map1}
  		\check{\mathbf{z}}_{k+1} = \boldsymbol{\Phi}\rb{\mathbf{z}_k,\mathbf{p}} \cdot \check{\mathbf{z}}_k
  	\end{equation}
  	to the ellipsoid
  	$\check{{\mathcal{E}}}_k$ in~(\ref{eq:thick_ell_center0}). The outer ellipsoid enclosure of the mapping (\ref{eq:map1}) is described by an ellipsoid with the shape matrix
  	\begin{equation}\label{eq:out_tmp}
  		\check{\mathbf{Q}}_{k+1} = \alpha_{k+1}^2  \cdot \boldsymbol{\Gamma}_{k+1} \cdot \boldsymbol{\Gamma}_{k+1}^T
  		\enspace, 
  	\end{equation}
  	where $\alpha_{k+1}\ge 0$ is the smallest value for which the LMI
  	\begin{equation}\label{eq:out}
  		\scalebox{0.9}{$
  			\mathbcal{{M}}_{k+1}:= \boldsymbol{\Lambda}
  			\begin{bmatrix}
  				- \mathbf{Q}_k^{-1} & \boldsymbol{\Phi}^T\rb{\mathbf{z}_k,\mathbf{p}} \cdot \tilde{\boldsymbol{\Phi}}^{-T} \\
  				\tilde{\boldsymbol{\Phi}}^{-1} \cdot {\boldsymbol{\Phi}}\rb{\mathbf{z}_k,\mathbf{p}} & -\alpha_{k+1}^2 \mathbcal{{R}}_{k}
  			\end{bmatrix} \boldsymbol{\Lambda}
  			\preceq 0
  			$}
  	\end{equation}
  	is satisfied for all $\mathbf{z}_k \in \square{\mathcal{E}_k}$ and $\mathbf{p} \in \sqb{\mathbf{p}}$ with the choice
  	\begin{equation}\label{eq:R_k_opt}
  		\mathbcal{R}_{k}:=\boldsymbol{\Gamma}_k \cdot \boldsymbol{\Gamma}_k^T\enspace .
  	\end{equation}
  	Alternatives for the choice of $\mathbcal{R}_{k}$ as well as advantages and disadvantages are discussed in \cite{a14030088}. In~(\ref{eq:out}), the symbol $\preceq$ denotes the negative semi-definiteness of the corresponding matrix expression and $\boldsymbol{\Lambda}$ is a preconditioning matrix chosen according to~\cite{Rauh_Jaulin_ICCAD}.
  	{A typical choice for the preconditioning is a block diagonal matrix $\boldsymbol{\Lambda}=\mathrm{blkdiag}\rb{\beta \mathbf{I}, \beta^{-1}\mathbf{I}}$ with $\mathbf{I}$ as the identity matrix and the square root $\beta= \sqrt{\min\{\lambda_i\rb{\mathbf{Q}_k}\}}$ of the smallest eigenvalue of $\mathbf{Q}_k$.} 
  	
  	\item[\textbf{P2}] Compute interval bounds for the term
  	\begin{equation}
  		\mathbf{b}_k = \rb{{\boldsymbol{\Phi}\rb{\mathbf{z}_k,\mathbf{p}}} - \tilde{\boldsymbol{\Phi}}} \cdot \boldsymbol{\mu}_k \in \sqb{\mathbf{b}_k}
  	\end{equation}
  	which accounts for a non-zero ellipsoid midpoint with $\mathbf{z}_k$, $\tilde{\boldsymbol{\Phi}}$, and $\mathbf{p}$ defined according to~(\ref{eq:thick_ell_complete}),~(\ref{eq:A_tilde}), and~(\ref{eq:par_box}). Inflate the ellipsoid bound described by the shape matrix~(\ref{eq:out_tmp}) according to \cite{Rauh_Jaulin}
  	\begin{align}
  		\label{eq:QO_kp1}
  		{\mathbf{Q}}_{k+1} & = \rb{1+\rho_{\mathrm{O},k+1}}^2 \cdot \check{\mathbf{Q}}_{k+1}~,\\
  		\rho_{\mathrm{O},k+1} & = \sup\left\{\left\|
  		\alpha_{k+1}^{-1} \cdot  \boldsymbol{\Gamma}_k^{-1} \cdot
  		\sqb{\mathbf{b}_k}
  		\right\|\right\} \label{eq:norm}
  		\enspace. 
  	\end{align}
  	\item[\textbf{P3}] Compute the updated ellipsoid midpoint 
  	\begin{equation}\label{eq:MP_new}
  		\boldsymbol{\mu}_{k+1} = 
  		\tilde{\boldsymbol{\Phi}} \cdot \boldsymbol{\mu}_k 
  	\end{equation}
  	and its factorized shape matrix 
  	\begin{equation}\label{eq:rho_gamma_quasi_lin}
  		\begin{split}
  			\boldsymbol{\Gamma}_{k+1}^\prime =
  			\alpha_{k+1} \cdot \rb{1+\rho_{\mathrm{O},k+1}}
  			\cdot
  			\tilde{\boldsymbol{\Phi}} \cdot \boldsymbol{\Gamma}_{k}^\prime
  			\, .
  		\end{split}
  	\end{equation}
  \end{description}

  \subsection{Stochastic Iterative-Learning Observer Design for Quasi-Linear State Equations}\label{sec:problem}
  To apply the previous set-based prediction step, initialized at $k=1$ with a suitable ellipsoidal state enclosure for forecasting a certain confidence set, we consider
  the quasi-linear discrete-time state-space representation
  \begin{equation}\label{eq:LTV2}
  	\begin{split}
  		\mathbf{x}_{k+1} & = 
  		\mathbf{A}\rb{{\mathbf{x}}_{k},\mathbf{p}} \cdot 
  		\mathbf{x}_{k}
  		+
  		\mathbf{E}\rb{{\mathbf{x}}_{k},\mathbf{p}}\cdot 
  		\mathbf{w}_{k} \\
  		\mathbf{y}_k & = \mathbf{C}\rb{{\mathbf{x}}_{k},\mathbf{p}} \cdot \mathbf{x}_{k} + \mathbf{v}_{k}
  	\end{split}
  \end{equation}
  with the state vector $\mathbf{x}_{k} \in \mathbb{R}^n$, the measured output vector $\mathbf{y}_{k} \in \mathbb{R}^m$ ($m\le n$), as well as the uncorrelated process and measurement noise vectors $\mathbf{w}_{k}\in \mathbb{R}^{n_\mathbf{w}}$ and $\mathbf{v}_{k}\in \mathbb{R}^{n_\mathbf{v}}$, respectively. Let both noise vectors be normally distributed with the covariances $\mathbf{C}_{\mathbf{w},k}$ and $\mathbf{C}_{\mathbf{v},k}$ and vanishing mean. The iterative-learning observer is designed in this section to estimate the state vector $\mathbf{x}_{k}$ as well as its uncertainty (expressed by its covariance) by a Kalman Filter-like procedure. This estimator does not only operate along the time domain~$k$ but also enhances the estimates successively from the trial $i$ to the trial $i+1$ {during which the underlying system operates along the same trajectories}. Here, the word trial means that the system repeats the same finite-duration control task multiple times and that the estimator is meant to enhance its accuracy each time the task is re-executed. To further allow for the identification of a systematic model mismatch, a lumped correction term $\boldsymbol{\delta}_{k}$ is added to the state equations~(\ref{eq:LTV2}) in the form
  \begin{equation}\label{eq:LTV_corr}
  	\mathbf{x}_{k+1} = 
  	\mathbf{A}\rb{{\mathbf{x}}_{k},\mathbf{p}} \cdot 
  	\mathbf{x}_{k}
  	+
  	\mathbf{E}\rb{{\mathbf{x}}_{k},\mathbf{p}}\cdot 
  	\mathbf{w}_{k}
  	+
  	\boldsymbol{\delta}_{k}
  	\enspace .
  \end{equation}
	Strategies for learning $\boldsymbol{\delta}_{k}$ during the execution of the iterative-learning estimator are described in \cite{Rauh_ECC_2022}.
  
  For the derivation of the learning-type framework, consider {two subsequent} trials $\xi=i$ and $\xi = i+1$ with the actually measured data $\mathbf{y}_{\mathrm{m},k}^\xi= \mathbf{C}_k^\xi \cdot \mathbf{x}_{k}^\xi + \mathbf{v}_{k}^\xi$  corresponding to the realizations of the general outputs $\mathbf{y}_{k}$ in~(\ref{eq:LTV2}) for these two trials.   
  
  For the recursive formulation of the iterative-learning observer, we assume that both process and measurement noise are uncorrelated and normally distributed with zero mean. The superscript $\mathrm{p}$ denotes the result of the prediction step, while the superscript $\mathrm{e}$ refers to the estimation result as the outcome of the measurement-based innovation step.
  
  \subsubsection{{Prediction Step}}
  
  {To perform a state prediction in the case of state- and parameter-dependent disturbance inputs $\mathbf{E}\rb{{\mathbf{x}}_{k},\mathbf{p}}$ according to~\cite{9841329}, define} an ellipsoid for the augmented state vector 
  \begin{equation}
  	\mathbf{z}_k = \begin{bmatrix}
  		\rb{\mathbf{x}_k^i}^T & \rb{\mathbf{x}_k^{i+1}}^T & \rb{\mathbf{w}_k^i}^T & \rb{\mathbf{w}_k^{i+1}}^T
  	\end{bmatrix}^T
  \end{equation}
  corresponding to the result of the preceding innovation step, augmented by the influence of the uncorrelated process noise in both trials $i$ and $i+1$ in the form
  \begin{equation}\label{eq:ell_estim1}
  	\mathcal{E}_k^{\mathrm{e},i|i+1}\rb{\begin{bmatrix}
  			\boldsymbol{\mu}_{k}^{\mathrm{e},i} \\
  			\boldsymbol{\mu}_{k}^{\mathrm{e},i+1} \\
  			\mathbf{0} \\
  			\mathbf{0}
  		\end{bmatrix}, \boldsymbol{\Gamma}_{k}^\mathrm{e}, r}
  \end{equation}
  with
  \begin{equation}
  	\boldsymbol{\Gamma}_{k}^\mathrm{e} = 
  	\begin{bmatrix}
  		\boldsymbol{\Gamma}_{k}^{\mathrm{e},i|i+1} & \mathbf{0}_{2n \times 2n_\mathbf{w}} \\
  		\mathbf{0}_{2 n_\mathbf{w} \times 2n} & 
  		\begin{bmatrix}
  			\mathbf{C}_{\mathbf{w},k} & \mathbf{0}\\
  			\mathbf{0} & \mathbf{C}_{\mathbf{w},k}
  		\end{bmatrix}^{\frac{1}{2}}
  	\end{bmatrix}
  \end{equation}
  containing the matrix square root $\boldsymbol{\Gamma}_{k}^{\mathrm{e},i|i+1}$ of the combined state covariance matrix
  \begin{equation}
  	\mathbf{C}_{k}^{\mathrm{e},i|i+1} = \boldsymbol{\Gamma}_{k}^{\mathrm{e},i|i+1} \cdot  \rb{\boldsymbol{\Gamma}_{k}^{\mathrm{e},i|i+1}}^T\enspace ,
  \end{equation}
  {in which joint distributions over the two subsequent trials are denoted by the superscript $i|i+1$,}
  as well as of the trial-independent noise covariance matrix $\mathbf{C}_{\mathbf{w},k}$, and the magnification factor $r\ge 1$ as a user-defined degree of freedom. {Here, in analogy to the Kalman Filter, the term \emph{innovation step} refers to the operation in which the predicted state information obtained from an open-loop evaluation of the system model into the future is corrected by taking into account measured data.}
  
  Then, the application of the system model {with the augmented dynamics matrix}
  \begin{equation}\label{eq:LTV2_ext}
  		\boldsymbol{\Phi}\rb{\mathbf{z}_k,\mathbf{p}}
  		= 
  		\begin{bmatrix}
  			\mathbf{A}\rb{{\mathbf{x}}_{k}^i,\mathbf{p}} & \mathbf{0}_{n \times n} &
  			\mathbf{E}\rb{{\mathbf{x}}_{k}^i,\mathbf{p}} & \mathbf{0}_{n \times n_\mathbf{w}}\\
  			\mathbf{0}_{n \times n} & \mathbf{A}\rb{{\mathbf{x}}_{k}^{i+1},\mathbf{p}} & 
  			\mathbf{0}_{n \times n_\mathbf{w}} & \mathbf{E}\rb{{\mathbf{x}}_{k}^{i+1},\mathbf{p}}\\
  			\mathbf{0}_{n_\mathbf{w} \times n} & \mathbf{0}_{n_\mathbf{w} \times n} &  \mathbf{I}_{n_\mathbf{w} \times n_\mathbf{w}} &  \mathbf{0}_{n_\mathbf{w} \times n_\mathbf{w}} \\
  			\mathbf{0}_{n_\mathbf{w} \times n} & \mathbf{0}_{n_\mathbf{w} \times n} &  \mathbf{0}_{n_\mathbf{w} \times n_\mathbf{w}} &  \mathbf{I}_{n_\mathbf{w} \times n_\mathbf{w}} \\
  		\end{bmatrix}
  \end{equation}
  to the ellipsoid~(\ref{eq:ell_estim1}) according to Sec.~\ref{sec:predict_ellipsoid} yields the ellipsoid
  \begin{equation}\label{eq:ell_pred1}
  	\mathcal{E}_{k+1}^{\mathrm{p},i|i+1}\rb{\begin{bmatrix}
  			\boldsymbol{\mu}_{k+1}^{\mathrm{p},i} \\
  			\boldsymbol{\mu}_{k+1}^{\mathrm{p},i+1} \\
  			\mathbf{0} \\
  			\mathbf{0}
  		\end{bmatrix}, \boldsymbol{\Gamma}_{k+1}^\mathrm{p}, r}
  	\enspace .
  \end{equation}
  {In this ellipsoid, the first $2n$ components $\begin{bmatrix}
  	\rb{\boldsymbol{\mu}_{k+1}^{\mathrm{p},i}}^T & 
  	\rb{\boldsymbol{\mu}_{k+1}^{\mathrm{p},i+1}}^T
  \end{bmatrix}^T$ of the midpoint vector represent the predicted expected value vector}
   and the covariance $\mathbf{C}_{k+1}^{\mathrm{p},i|i+1}$ is obtained by extracting the upper left $(2n \times 2n)$ block of the matrix product $\boldsymbol{\Gamma}_{k+1}^{\mathrm{p}} \cdot  \rb{\boldsymbol{\Gamma}_{k+1}^{\mathrm{p}}}^T$.

  \subsubsection{Innovation Step}\label{sec:innov}
  The measurement-based innovation step makes use of the deviations
  \begin{align}\label{eq:del_y1}
  	\Delta \mathbf{y}_k^i & = 
  	{\mathbf{y}_{\mathrm{m},k}^i - \mathbf{C}_k^i \cdot \boldsymbol{\mu}_{k}^{\mathrm{p},i}}
	 \intertext{and}
  	\Delta \mathbf{y}_k^{i+1} & = 
  	{\mathbf{y}_{\mathrm{m},k}^{i+1} - \mathbf{C}_k^{i+1} \cdot \boldsymbol{\mu}_{k}^{\mathrm{p},i+1}}\label{eq:del_y2}
  \end{align}
  between the measured data in the trials $i$ and $i+1$, respectively,  and the corresponding output forecasts based on the prediction step of the previous subsection. Using these output deviations, and assuming the standard detectability requirements of the Kalman Filter design to be satisfied,
  the expected values are updated according to~\cite{Rauh_ECC_2022} by
  \begin{equation}\label{eq:innov1}
  	\begin{bmatrix}
  		\boldsymbol{\mu}_{k}^{\mathrm{e},i} \\
  		\boldsymbol{\mu}_{k}^{\mathrm{e},i+1}
  	\end{bmatrix} = \begin{bmatrix}
  		\boldsymbol{\mu}_{k}^{\mathrm{p},i} \\
  		\boldsymbol{\mu}_{k}^{\mathrm{p},i+1}
  	\end{bmatrix} + 
  	\tilde{\mathbf{H}}_k
  	\cdot 
  	\begin{bmatrix}
  		\mathbf{y}_{\mathrm{m},k}^i \\
  		\mathbf{y}_{\mathrm{m},k}^{i+1}
  	\end{bmatrix} -
  	\tilde{\mathbf{H}}_k
  	\tilde{\mathbf{C}}_k
  	\cdot 
  	\begin{bmatrix}
  		\boldsymbol{\mu}_{k}^{\mathrm{p},i} \\
  		\boldsymbol{\mu}_{k}^{\mathrm{p},i+1}
  	\end{bmatrix} \enspace ,
  \end{equation}
  where the combined output matrix 
  \begin{equation}
  	\tilde{\mathbf{C}}_k :=
  	\begin{bmatrix}
  		\mathbf{C}_k^i & \mathbf{0} \\
  		\mathbf{0} & \mathbf{C}_k^{i+1}
  	\end{bmatrix}
  \end{equation}
  results from pointwise evaluations of the quasi-linear system's output matrix according to 
  \begin{align}
  \mathbf{C}_k^i &:= \mathbf{C}_k\rb{\boldsymbol{\mu}_{k}^{\mathrm{p},i},\mathrm{mid}\rb{\sqb{\mathbf{p}}}}
  \intertext{and} 
  \mathbf{C}_k^{i+1} &:= \mathbf{C}_k\rb{\boldsymbol{\mu}_{k}^{\mathrm{p},i+1},\mathrm{mid}\rb{\sqb{\mathbf{p}}}}
  \enspace .
  \end{align}
  
  {To optimize the estimator performance, the filter gain matrix $\tilde{\mathbf{H}}_k$ needs to be tuned. This is possible by a minimization of the uncertainty on the estimated state variables. Inspired by the derivation of the Kalman Filter \cite{Kalman:1960}, a minimization of the estimation error covariance is performed. For that purpose, this covariance is computed according to}
  \begin{align}
  	\mathbf{C}_{k}^{\mathrm{e},i|i+1} &= 
  	\mathrm{E}\left\{
  	\begin{bmatrix}
  		\mathbf{x}_k^i - \boldsymbol{\mu}_k^{\mathrm{e},i} \\
  		\mathbf{x}_k^{i+1} - \boldsymbol{\mu}_k^{\mathrm{e},i+1}
  	\end{bmatrix} \cdot 
  	\begin{bmatrix}
  		\mathbf{x}_k^i - \boldsymbol{\mu}_k^{\mathrm{e},i} \\
  		\mathbf{x}_k^{i+1} - \boldsymbol{\mu}_k^{\mathrm{e},i+1}
  	\end{bmatrix}^T
  	\right\} \nonumber \\
  	& = 
  	\mathrm{Cov}\left\{
  	\begin{bmatrix}
  		\mathbf{x}_k^i - \boldsymbol{\mu}_k^{\mathrm{e},i} \\
  		\mathbf{x}_k^{i+1} - \boldsymbol{\mu}_k^{\mathrm{e},i+1}
  	\end{bmatrix}
  	\right\} \nonumber \\ 
  	& = \mathbf{M}_k  \mathbf{C}_{k}^{\mathrm{p},i|i+1}  \mathbf{M}_k^T + \tilde{\mathbf{H}}_k  \tilde{\mathbf{C}}_{\mathbf{v},k}  \tilde{\mathbf{H}}_k^T\enspace ,
  	\label{eq:err_cov1}
  \end{align}
  where
  \begin{equation}
  	\mathbf{M}_k = \begin{bmatrix}
  		\mathbf{I} & \mathbf{0} \\
  		\mathbf{0} & \mathbf{I}
  	\end{bmatrix} -
  	\tilde{\mathbf{H}}_k \tilde{\mathbf{C}}_k
  	\enspace .
  \end{equation}
  {To ensure causality, i.e., the update in iteration $i$ does not depend on the future iteration $i+1$, the gain matrix is structured according to}
  \begin{equation}
  	\tilde{\mathbf{H}}_k :=
  	\begin{bmatrix}
  		\mathbf{H}_{1,k}^{i+1} & \mathbf{0} \\
  		\mathbf{H}_{2,k}^{i+1} & \mathbf{H}_{1,k}^{i+1}-\mathbf{H}_{2,k}^{i+1}
  	\end{bmatrix}
  	\enspace .
  \end{equation}
  
  As shown in~\cite{Rauh_ECC_2022}, {the minimization task described above can be solved by differentiating the trace of the covariance $\mathbf{C}_{k}^{\mathrm{e},i|i+1}$ with respect to the gain matrix and setting the resulting expression, being linear in both gain matrices  $\mathbf{H}_{1,k}^{i+1}$ and $\mathbf{H}_{2,k}^{i+1}$, to zero. A relevant intermediate step is to determine the predicted covariance that is} partitioned in a blockwise manner according to
  \begin{equation}
  	\mathbf{C}_{k}^{\mathrm{p},i|i+1} = \begin{bmatrix}
  		\mathbf{C}_{\mathrm{A},k}^{\mathrm{p}} & \mathbf{C}_{\mathrm{B},k}^{\mathrm{p}} \\
  		{\rb{\mathbf{C}_{\mathrm{B},k}^{\mathrm{p}}}^T} & \mathbf{C}_{\mathrm{C},k}^{\mathrm{p}}
  	\end{bmatrix}
  	\enspace .
  \end{equation}
Moreover, the residual covariance is defined as
  \begin{equation}\label{eq:resid}
  	\mathbf{P} \cdot 
  	{\tilde{\boldsymbol{\Gamma}}_{k}^{\mathrm{e}} \cdot \rb{\tilde{\boldsymbol{\Gamma}}_{k}^{\mathrm{e}}}^T} \cdot \mathbf{P}^T
  	+\tilde{\mathbf{C}}_{\mathbf{v},k} = 
  	\begin{bmatrix}
  		\mathbf{C}_{\mathrm{A},k} & \mathbf{C}_{\mathrm{B},k} \\
  		{\mathbf{C}_{\mathrm{B},k}^{T}} & \mathbf{C}_{\mathrm{C},k}
  	\end{bmatrix}
  \end{equation}
  with the projection matrix 
  $
  \mathbf{P} =
  \begin{bmatrix}
  	\mathbf{I}_{m\times m} & \mathbf{0}_{m\times (n-m)}
  \end{bmatrix}
  $, $\tilde{\boldsymbol{\Gamma}}_{k}^{\mathrm{e}}$ defined in the following,
  and the trial-independent measurement noise covariance matrix
  \begin{equation}
  	\tilde{\mathbf{C}}_{\mathbf{v},k} := \begin{bmatrix}
  		{\mathbf{C}}_{\mathbf{v},k} & \mathbf{0} \\
  		\mathbf{0} & {\mathbf{C}}_{\mathbf{v},k}
  	\end{bmatrix}
  	\enspace .
  \end{equation}
  
  As a specific feature introduced in \cite{9841329}, aiming at a robustification of the innovation stage against nonlinearities, the matrix $\tilde{\boldsymbol{\Gamma}}_{k}^{\mathrm{e}}$ is obtained by the consideration of the quasi-linearity of the output equation with the help of the ellipsoid 
  \begin{equation}\label{eq:ell_pred2_k}
  	\mathcal{E}_{k}^{\mathrm{p},i|i+1}\rb{\begin{bmatrix}
  			\boldsymbol{\mu}_{k}^{\mathrm{p},i} \\
  			\boldsymbol{\mu}_{k}^{\mathrm{p},i+1} 
  		\end{bmatrix}, \boldsymbol{\Gamma}_{k}^{\mathrm{p},i|i+1}, r}
  \end{equation}
  that is propagated through a quasi-linear system model in the form~(\ref{eq:sys_QL}) with the associated system matrix ($\mathbf{y}_{\mathrm{m},k}^{i}, \mathbf{y}_{\mathrm{m},k}^{i+1} \in \mathbb{R}^m$)
  \begin{equation}\label{eq:LTV2b}
\boldsymbol{\Phi}\rb{\mathbf{z}_k,\mathbf{p}}
  		= \begin{bmatrix}
  				\mathbf{C}\rb{{\mathbf{x}}_{k}^{i},\mathbf{p}}  & \mathbf{0}_{m\times n} \\
  				\mathbf{0}_{m\times n}  & \mathbf{C}\rb{{\mathbf{x}}_{k}^{i+1},\mathbf{p}}  \\
  				\begin{matrix}
  					\mathbf{0}_{(n-m)\times m} & \mathbf{I}_{(n-m) \times (n-m)}
  				\end{matrix} & 
  				\mathbf{0}_{(n-m)\times n} 
  				\\
  				\mathbf{0}_{(n-m)\times n} &\begin{matrix}
  					\mathbf{0}_{(n-m)\times m} & \mathbf{I}_{(n-m) \times (n-m)}
  				\end{matrix}
  			\end{bmatrix}
  			\enspace ,
  \end{equation}
  where 
  $\boldsymbol{\Phi}\rb{\mathbf{z}_k,\mathbf{p}}\in \mathbf{R}^{{2n}\times 2n}$. The evaluation of this quasi-linear model yields an ellipsoid 
  \begin{equation}\label{eq:ell_pred2_ka}
  	\mathcal{E}_{k}^{\mathrm{e}}\rb{
  		\tilde{\boldsymbol{\mu}}_{k}^{\mathrm{e}} 
  		, \tilde{\boldsymbol{\Gamma}}_{k}^{\mathrm{e}}, r}
  \end{equation}
  from which the associated shape matrix ${\tilde{\boldsymbol{\Gamma}}_{k}^{\mathrm{e}} \cdot \rb{\tilde{\boldsymbol{\Gamma}}_{k}^{\mathrm{e}}}^T}$ is computed in Eq.~(\ref{eq:resid}), followed by an extraction of the upper left ($2m \times 2m$) block that corresponds to the actually measured output quantities. This subblock extraction is performed by the multiplication with the projection matrix $\mathbf{P}$ in the first summand of Eq.~(\ref{eq:resid}).

  {As shown in further detail in~\cite{Rauh_ECC_2022}, the optimal iterative-learning observer gains are given by}
  	\begin{equation}\label{eq:gains}
  		\begin{split}
  			\left[
  			\begin{array}{c;{2pt/2pt}c}
  				\mathbf{H}_{1,k}^{i+1} & \mathbf{H}_{2,k}^{i+1}
  			\end{array} 
  			\right]& = 
  			\left[
  			\begin{array}{c;{2pt/2pt}c}\!\!\!
  				\rb{{\mathbf{C}_k^{i}}\mathbf{C}_{\mathrm{A},k}^\mathrm{p}+
  					{\mathbf{C}_k^{i+1}} \mathbf{C}_{\mathrm{C},k}^\mathrm{p}}^T \!\!
  				& \!\!\rb{{\mathbf{C}_k^{i}} {\mathbf{C}_{\mathrm{B},k}^\mathrm{p}}
  					-{\mathbf{C}_k^{i+1}} {\mathbf{C}_{\mathrm{C},k}^\mathrm{p}} }^T\!\!\!
  			\end{array} 
  			\right] \\
  			& \qquad \phantom{=}\cdot \begin{bmatrix}
  				\mathbf{C}_{\mathrm{A},k}+\mathbf{C}_{\mathrm{C},k} & \star \\
  				\mathbf{C}_{\mathrm{B},k}-\mathbf{C}_{\mathrm{C},k} & \mathbf{C}_{\mathrm{A},k} - \rb{\mathbf{C}_{\mathrm{B},k}+\mathbf{C}_{\mathrm{B},k}^T}+\mathbf{C}_{\mathrm{C},k}
  			\end{bmatrix}^{-1}
  			\!\!\!\!.
  		\end{split}
  	\end{equation}
  	{It should be noted at this point that setting all coupling terms between the iterations $i$ and $i+1$ to zero leads exactly to the same gain matrix that would be obtained when designing a Kalman filter if all matrices in~(\ref{eq:LTV_corr}) are state-independent and only depend on precisely known parameters.}
  
\section{Conclusions and Future Work}\label{sec:concl}

In this paper, a comprehensive overview of recent advances concerning the offline and online use of interval and other set-based approaches for control and state estimation has been given. We have equally addressed methodological approaches and selected fields of practical applications in our {selection}, where we have put a focus on the tasks of parameter estimation, robust and model predictive control, as well as state estimation procedures.

Future work will focus especially on widening the field of applications by an investigation of control and state estimation tasks for distributed and large-scale, interconnected systems with parameter uncertainty and disturbances that are, among others, omnipresent in control scenarios in the Industry~4.0 as well as in the domain of energy systems. To cope with such systems, we will focus especially on the development of decentralized techniques for the parameter identification of large-scale system models, which will again form the basis for control and state estimator design. Moreover, we will further develop approaches that allow for removing restrictive monotonicity assumptions as already shown in works such as \cite{Wang.2018,10057091,9867741}.
  
\bibliographystyle{alphaurl}
\bibliography{literatur}

\end{document}